\documentclass[a4paper,twocolumn,twoside,fleqn,10pt]{article}
\usepackage{amsmath}
\usepackage{authblk}
\usepackage[svgnames]{xcolor}
\definecolor{blue}{RGB}{0,0,255}
\usepackage{caption}
\usepackage[font={color=blue},figurename=Fig.,labelfont={it}]{caption}
\usepackage{bm}
\usepackage{subcaption}
\usepackage[varg]{txfonts}
\usepackage{graphicx}
\usepackage{pdfpages}
\usepackage{wasysym}
\usepackage{rotating}
\usepackage{hyperref}
\usepackage{indentfirst}
\usepackage{float}
\usepackage{lipsum} 
\usepackage{setspace} 
\usepackage{indentfirst}
\usepackage[english]{babel}
\usepackage[switch,columnwise]{lineno}
\usepackage[left=2cm,right=2cm,top=2cm,bottom=2cm]{geometry}

\hypersetup{
    colorlinks=true,                          
    linkcolor=blue, 
    citecolor=blue, 
    urlcolor=blue,
    linktoc=all
}

\newcommand{\blue}[1]{\textcolor{blue}{#1}}

\newcommand{\etc}{\textit{etc }\ldots }
\newcommand{\eg}{\textit{eg }}
\newcommand{\cf}{\textit{cf }}
\newcommand{\ie}{\textit{ie }}

\newcommand{\degree}{$^{\circ}$}
\newcommand{\Poisson}{\blue{Poisson} }
\newcommand{\Gauss}{\blue{Gauss} }
\newcommand{\Legendre}{\blue{Legendre} }
\newcommand{\Lagrange}{\blue{Lagrange} }
\newcommand{\Laplace}{\blue{Laplace} }
\newcommand{\LeMouel}{\blue{Le Mouël} }

\usepackage{fancyhdr}
\begin{document}
\title{Is the Earth’s magnetic field a constant ? a legacy of Poisson}
\author[1]{J-L. Le Mouël}
\author[1]{F. Lopes\thanks{lopesf@ipgp.fr}}
\author[1]{V. Courtillot}
\author[2]{D. Giber}
\author[3]{J-B. Boulé}
			
\affil[1]{Universit\'e Paris Cité, Institut de Physique du globe de Paris, CNRS UMR 7154, F-75005 Paris, France}
\affil[2]{LGL-TPE - Laboratoire de Géologie de Lyon - Terre, Planètes, Environnement, Lyon, France}
\affil[3]{CNRS UMR7196, INSERM U1154, Museum National d'Histoire Naturelle, Paris, F-75005,  France}

\date{\today}
\maketitle

\abstract {In the report he submitted to the Académie des Sciences, Poisson imagined a set of concentric spheres at the origin of the Earth's magnetic field. It may come as a surprise to many that Poisson as well as Gauss both considered the magnetic field to be constant. We propose in this study to test this surprising assertion for the first time evoked by Poisson (1826). First, we will present a development of Maxwell's equations in the framework of a static electric field and a static magnetic field in order to draw the necessary consequences for the Poisson hypothesis. In a second step, we will see if the observations can be in agreement with Poisson (1826). To do so, we have chosen to compare 1) the polar motion drift and the secular variation of the Earth's magnetic field, 2) the seasonal pseudo-cycles of day length together with those of the sea level recorded by different tide gauges around the globe and those of the Earth's magnetic field recorded in different magnetic observatories. We then propose a mechanism, in the spirit of Poisson, to explain the presence of the 11-year in the magnetic field. We test this mechanism with observations and finally we study closely the evolution of the g10 coefficient of the IGFR over time.}

%
\tableofcontents

\section{\label{sec1} Introduction}
	The birth of geomagnetism as a science can be dated as August 8, 1269: on that day \blue{Petrus Peregrinus} wrote three letters (\cite{Petrus1269}; see \textit{f.i.} \blue{Courtillot and Le Mouël} \cite{Courtillot2007}) during the siege of the city of Lucera in the Italian region of Puglie. The letters can be considered as the first scientific article on geomagnetism, 331 years before the famous De Magnete by \blue{Gilbert} (\cite{Gilbert1600}). Expressed in modern terms, \blue{Peregrinus} wrote that the Earth had a magnetic field with a dipolar structure and that some rocks or minerals were magnetized. One could carve a sphere out of magnetite, pierce a hole through its center and it would oscillate around the northward direction, a thought experiment that announced the compass.
	
	In the three centuries that unfurled since then, a lot has been discovered: the variation of inclination with latitude, the daily, annual and other periodic variations (\cite{Mayaud1972,Courtillot1977,LeMouel2019a}), irregular variations such as events linked to solar activity, the secular variation and its sudden jerks (\cite{Ducruix1980,LeMouel1981,Courtillot1984}). In order to handle an ever increasing data base, magnetic indices (\eg aa, Dst, Kp, \etc) were introduced (\eg \cite{Bartels1932,Bartels1939,Chapman1940,Mayaud1968,Mayaud1971,Mayaud1972,Mayaud1978,Mayaud1980,LeMouel2004,LeMouel2019a}). It was found that quasi periodical variations of the magnetic field (\cite{Mayaud1972,Courtillot1977,LeMouel2019a}), and also sun spots (\cite{LeMouel2020a}) followed a \cite{Kolmogorov1941} power law with exponent -5/3.
	
	In a series of papers that started with \blue{Le Mouël} in 1984 (\cite{LeMouel1984}) and continued with \blue{Jault et \textit{al.}} (\cite{Jault1988}), and \blue{Jault and Le Mouël} (\cite{Jault1989,Jault1990,Jault1991}), these authors found that the trends of (1) magnetic secular variation, (2) polar motion and (3) length of day were strongly correlated. They proposed to explain these observations with a coupling mechanism in which flow in a cylinder tangent to the core and the rotation axis exchange torques at the core-mantle boundary. The solution for flow on the cylinder (see \cite{Jault1991}, system 6) is the same as that generated by an internal gravitational wave in a rotating fluid (\eg \cite{Landau1987}) also known as \blue{Proudman} (\cite{Proudman1916}) flow. But this mechanism encountered serious difficulties with the orders of magnitude of physical parameters such as CMB topography. Also, the torque exerted by the fluid pressure and the electromagnetic torque were too weak to validate the model, as concluded by \blue{Jault and Le Mouël} (\cite{Jault1991}).

	Actually, the model of a cylinder tangent to the core is very close in spirit to that envisioned by \blue{Poisson} (\cite{Poisson1826}). In the report he submitted to the Académie des Sciences, prior to an oral communication, \blue{Poisson} imagined a set of concentric spheres in place of a cylinder. \blue{Poisson} (\cite{Poisson1826}) was the first scientist to describe the magnetic field as a series of spherical harmonics, a decade before \blue{Gauss} did (\cite{Gauss1837}), Part 5, chapter 1, ‘Allgemeine Theorie des Erdmagnetismus’). He also invented a technique to measure the absolute value of the horizontal component of the magnetic field (\cite{Poisson1825,Lepretre2005}), seven years before \blue{Gauss} (\cite{Gauss1832}) did.

	It may come as a surprise to many that \blue{Poisson} (\cite{Poisson1826}) as well as \blue{Gauss} (\cite{Gauss1837} both considered the magnetic field to be constant. This was an axiomatic basis for the development into spherical harmonics. There are very clear statements to this effect in the writings of both scientists. In \blue{Poisson} (\cite{Poisson1826}) page  49, one reads (our translation from the French): "\textit{We will assume that the hollow sphere be magnetized under the influence of a force that be the same in magnitude and direction for all its points, such as the magnetic action of Earth, for instance}". And in page 54:"\textit{Since time does not enter these formulae, a consequence is that, after the first instants of rotation, that we did not mention, the action of the rotating sphere on a given point will be constant in magnitude and in direction}". \blue{Gauss} (\cite{Gauss1837}), part 5, chapter 1, paragraph 2, page 6, \blue{Gauss} writes (our translation from the German): "\textit{$\ldots$ magnetism consists only in galvanic currents (that is constant currents) that persist in the smallest parts of the bodies $\ldots$}". Gauss develops his theory very quickly (pages 18 to 23, paragraphs 14 to 27), without any physical proof. And his mathematical proof is exactly that found in \blue{Legendre} (\cite{Legendre1785}) or \blue{Laplace}(\cite{Laplace1799}) for the gravitational field. In contrast, the 130 pages of \blue{Poisson} (\cite{Poisson1826}) memoir are devoted both to the full physical and mathematical proofs of the magnetic field description.

	Given that \Poisson (\cite{Poisson1826}) work precedes and is more complete than that of \Gauss (\cite{Gauss1837}), it is only fair to recognize that the former was the first to develop the magnetic field in spherical harmonics and to state that this magnetic field was constant.
	
	We do not propose to follow, as \Poisson (\cite{Poisson1826}) did, the description of a magnetic field based on \blue{Maupertuis} (\cite{Maupertuis1750}) principle of least action, but rather to pursue our previous presentation of the laws of gravitation \blue{Lopes et al.} (\cite{Lopes2023}) following \blue{Lagrange} (\cite{Lagrange1788}) and extend it to the case (and consequences) of the constancy of the electric and magnetic fields, giving their full physical meaning to the moments. This is the aim of section \ref{sec2}. In section \ref{sec3} we try to explain the variations that one can measure within the frame of \Poisson (\cite{Poisson1826}) paradigm. We confront the previous theoretical developments with modern observations in section \ref{sec4} and conclude in section \ref{sec5}.

\section{On the constancy of the magnetic field}\label{sec2}
	\subsection{Some consequences of Maxwell’s equations}\label{sec2-1}
	One can learn a lot from a Lagrangian approach to the derivation of Maxwell’s equations. Following \blue{Maupertuis} (\cite{Maupertuis1750}), one only needs to know the action of a moving charged particle in an electromagnetic (\textbf{EM}) field, associated with the action of its interaction with that field, to derive the first pair of equations :
\begin{subequations}	
	\begin{align}
		&\textrm{rot} \textbf{E} = -\dfrac{1}{c} \dfrac{\partial \textbf{H}}{\partial t}\label{eq:01a}\\
		&\textrm{div} \textbf{H} = 0 \label{eq:01b}
	\end{align}
	Adding the action of field \textbf{EM}, one obtains the second pair:

	\begin{align}
		&\textrm{rot} \textbf{H} = \dfrac{1}{c} \dfrac{\partial \textbf{E}}{\partial t} + \dfrac{4\pi}{c} \textbf{j} 		\label{eq:01c} \\
		&\textrm{div} \textbf{E} = 4\pi \rho 		\label{eq:01d}
	\end{align}
\end{subequations}

		Equations (\ref{eq:01a}) to (\ref{eq:01d}) link in a symmetrical way the magnetic field (\textbf{H}) to the electric field (\textbf{E}). $c$ is the velocity of the charged particle $\rho$ associated to current density $\textbf{j}$. It is important to note that, without knowing the action of \textbf{EM}, one has access to important properties : the space component, which is actually the field H, is conserved (\ref{eq:01b}). By analogy to Euler’s (continuity\footnote{The continuity equation from fluid mechanics}) equation, "magnetic charges" do not exist. Equation (\ref{eq:01a}) implies that as soon as \textbf{H} varies with time, a field \textbf{E} that is perpendicular (rot) and in quadrature with \textbf{H} is created (instantly hence the term $1/c$). The second pair of equations is fully symmetrical. Equation (\ref{eq:01d}) implies that there exists an "electric charge" ($\rho$) that locally deforms the field \textbf{E}. In a vacuum (\ref{eq:01d}) has exactly the same physical meaning as (\ref{eq:01b}). But one must add a current density to the time variation of \textbf{E} to propagate the field \textbf{H} (\ref{eq:01c}). A side note on (\ref{eq:01c}) also known as Maxwell-Ampere equation. The classical understanding is that magnetic fields can be generated in two different ways, either by electrical currents (Ampere’s theorem), or by time changes of field \textbf{E}, or the sum of both. The Lagrangian approach clarifies the picture. An \textbf{EM} field is defined by its 4-vector potential $A_i(=\varphi,\textbf{A})$, where $\varphi$ is the time component (called the scalar potential, linked to \textbf{E}) and $\textbf{A}$ the space component (called the vector potential and linked to \textbf{H}). Charges that move in the field must obey the same decomposition ; one therefore introduces a 4-vector current density $j^i(=c\rho,\textbf{j})$, with a scalar charge density ($\rho$) found in (\ref{eq:01d}) and a vector current density ($\textbf{j}$).
		
		 The \textbf{EM} field described by the Maxwell equations must be in one of the three following forms: electrostatic, magnetostatic or a propagating field (wave propagation) that will not be discussed further in this paper. 
	 
	\subsection{The electrostatic field}\label{sec2-2}
	Equations (\ref{eq:01a}) and (\ref{eq:01d}) reduce to:
\begin{subequations}	
	\begin{align}
		&\textrm{div} \textbf{E} = 4\pi \rho \label{eq:02a} \\
		&\textrm{rot} \textbf{E} = 0 		\label{eq:02b} 
	\end{align}
	
	\textbf{E} derives from scalar potential ($\varphi$): 
	\begin{center}
	 	$\textbf{E} = - \textrm{grad}\ \varphi$
 	\end{center}
	
	leading to the Poisson equation:
	\begin{equation}
		\textrm{div} (\textrm{grad})\ \varphi = \Delta \varphi = -4\pi \rho \label{eq:02c}
 	\end{equation}		
	
	In a vacuum ($\rho = 0$), the scalar potential verifies the Laplace equation:
	\begin{center}
		$\Delta\ \varphi = 0$
	\end{center}

	The field produced by a point charge ($e$) will be directed along the vector having the charge as one of its extremities. \textbf{E} is a radial field. The absolute value of \textbf{E} depends only on the distance $R$ to $e$. Applying the divergence theorem to (\ref{eq:02a}):	
	\begin{center}
	$\textrm{div} \textbf{E} \longmapsto \iiint_{V} \textrm{div}  \textbf{E}\  dV = \iint_ {S} \textbf{E} . d\textbf{S}$
		\end{center}
		
		The flux of \textbf{E} across a spherical surface with radius $R$ centered on $e$ is $4\pi R^2 \textbf{E}$ and also equals $4\pi e$ from Gauss’s theorem, 
	\begin{center}
		$\oint \textbf{E} d\textbf{f} = 4\pi \int e dV$. 
	\end{center}			
	
	Finally, in vector form:
	\begin{equation}
		\textbf{E} = \dfrac{e\textbf{R}}{R^3}				\label{eq:02d}
	\end{equation}			
	
	Thus, the field produced by a point charge is inversely proportional to the square of the distance to $e$ (Coulomb’s law). The potential associated to this field is:
	\begin{equation}
			\varphi = \dfrac{e}{\textbf{R}}			\label{eq:02e}
	\end{equation}
	
	and for a system of charges:
	\begin{equation}
		\varphi = \sum_{i} \dfrac{e_i}{\textbf{R}_i}		\label{eq:02f}
	\end{equation}
\end{subequations}	
	
	Let us observe this field (\ref{eq:02f}) at a distance that is large compared to the charge system’s dimension, that is far enough so that their relative motions can be considered as constant (in the sense of Lagrangian mechanics). This allows one to use the concept of "moment". 
	
\begin{subequations}	
	
	Let us choose a coordinate system whose origin lies within the charge system and let $r_i$ be their respective vector radii. The total observed potential at point $R_o$ is:
	\begin{equation}
		\varphi = \sum \dfrac{e_a}{|\textbf{R}_o-\textbf{r}_i|}	\label{eq:03a}
	\end{equation}
	
	For $R_o \gg r_i$ and thanks to the generalized Maclaurin expansion given by \cite{Legendre1785}, we can develop (\ref{eq:03a}) in a series of powers of $\dfrac{r_i}{R_o}$, using the first order formula: 
	\begin{equation*}
	f(\textbf{R}_o-\textbf{r}) \approx f(\textbf{R}_o) - \textbf{r}\ \textrm{grad} f(\textbf{R}_o)
	\end{equation*}
	
	thus (\ref{eq:03a}) becomes: 
	\begin{equation}	
		\varphi = \dfrac{\sum e_i}{\textbf{R}_o} - \sum e_{i}\textbf{r}_{i} \textrm{grad} \dfrac{1}{\textbf{R}_o}
	\label{eq:03b}
	\end{equation}

	The sum $\textbf{d} = \sum e_{i} \textbf{r}_i$ is the dipolar moment of the charge system. By analogy with a system of masses, this dipolar moment is the mathematical equivalent to the tensor of moments of inertia of order 2 in geodesy. If the sum of charges is zero, the dipolar moment does not depend on the choice of the origin ($\textbf{r}^{‘} = \textbf{r} + \textbf{k}$), $k$ constant, $\sum e_i = 0$, then $\textbf{d}^{'} = \sum e_i \textbf{r}_{i}^{'} = \sum e_i \textbf{r}_{i} + \textbf{k}  \sum e_i =   \textbf{d}$) .
	
	The potential at large distances can be written:
	\begin{equation}	
		\varphi = -\textbf{d}\nabla \dfrac{1}{\textbf{R}_o} = \dfrac{\textbf{d} \textbf{R}_o}{R_o^3}
		\label{eq:03c}
	\end{equation}
	
	and:	
\begin{equation*}		
	\begin{split}
		\textbf{E} &= - \textrm{grad} \varphi \quad (=(\textbf{d}\nabla) \nabla \dfrac{1}{\textbf{R}_o})\\
	&= - \textrm{grad} \dfrac{\textbf{d}\textbf{R}_o}{R_o^3} = -\dfrac{1}{R_o^3}\textrm{grad} (\textbf{d}\textbf{R}_o) - (\textbf{d}\textbf{R}_o) \textrm{grad} \dfrac{1}{R_o^3}\\
    &= -\dfrac{\textbf{d}}{R_o^3} \nabla \textbf{R}_{o} - \dfrac{\textbf{R}_{o} }{R_o^3}\nabla \textbf{d}-\textbf{dR}_{o}\nabla \dfrac{1}{R_o^3}\\
    &= -\dfrac{\textbf{d}}{R_o^3} - 0 + \dfrac{3\textbf{dR}_o}{R_o^3}\\
    &= \dfrac{3(\textbf{nd})\textbf{n} - \textbf{d}}{R_o^3}
	\end{split}
\end{equation*}

	where \textbf{n} is the unit vector oriented towards $R_o$. At large distances, the potential is inversely proportional to the square of distance and \textbf{E} to its cube. \textbf{E} is axially symmetrical about \textbf{d}. In a plane where the direction of \textbf{d} is that of the $z$ axis,  the Cartesian components of \textbf{E} are:
	\begin{equation}
		E_z = d.\ \dfrac{3 \textrm{cos}^{2} \theta -1}{R_o^3}, \quad E_z = d.\ \dfrac{3 \textrm{sin}\theta \textrm{cos}\theta }{R_o^3}
		\label{eq:03d}
	\end{equation}

	and the radial and tangential components are: 
	\begin{equation}
	E_r = d.\ \dfrac{2 \textrm{cos}\theta}{R_o^3}, \quad E_{\theta} = -d.\ \dfrac{\textrm{sin}\theta }{R_o^3}
			\label{eq:03e}
	\end{equation}
\end{subequations}	

	$\theta$ being the angle between $z$ and $\textbf{R}_o$. We note that equations (\ref{eq:03d}) and (\ref{eq:03e}) are the same as those of the components of a dipolar magnetic field with d being replaced by $\dfrac{\mu_0}{4\pi}$(\eg \LeMouel in \cite{Coulomb1973a}, chapter 26, page 40, system 24).
	
	As done by \cite{Laplace1799}, one can always develop the scalar potential $\varphi$ as a sum of contributions in ascending powers of $\dfrac{1}{\textbf{R}_0}$, the term $\varphi^{(n)}$ being proportional to $\dfrac{1}{\textbf{R}_o^{n+1}}$,
\begin{subequations}	
	\begin{equation}
		\varphi = \varphi^{(0)}+ \varphi^{(1)} + \varphi^{(2)} + \ldots
		\label{eq:04a}
	\end{equation}

	The first term $\varphi^{(0)}$ is determined by the sum of all charges or masses for \Laplace (\cite{Laplace1799}), for whom this term can never be zero. As we have just seen, when the sum of electric charges is zero, one is led to the electrostatic components (\ref{eq:03d}) and (\ref{eq:03e}). The second term, $\varphi^{(1)}$ is the dipolar one, determined by its dipolar moment d. One can continue the development in a \Legendre (\cite{Legendre1785}) series. The next term would be: 
	\begin{equation}
		\varphi^{(2)} = \dfrac{1}{2} \sum e x_i x_j \dfrac{\partial^2 }{\partial X_i \partial X_j} \dfrac{1}{R_o}	
		\label{eq:04b}
	\end{equation}	
	
	where the $x$ coordinates are the components of $\textbf{r}$ and $X$ those of $\textbf{R}_o$. We note that:
	\begin{equation*}	
		\Delta \dfrac{1}{R_o} \equiv \delta _{ij} \dfrac{\partial^2 }{\partial X_i \partial X_j} \dfrac{1}{R_o} = 0
	\end{equation*}	

One can then write (\ref{eq:04b}) as:
	\begin{equation*}	
		\varphi^{(2)} = \dfrac{1}{2} \sum e (x_i x_j -\dfrac{1}{3} r^2 \delta_{ij})  \dfrac{\partial^2 }{\partial X_i \partial X_j} \dfrac{1}{R_o}
	\end{equation*}	
	
	The tensor $D_{ij} = \sum e (3x_i x_j – r^2 \delta_{ij})$ is the quadrupolar moment of the charge system. Thus:
	\begin{equation}	
		\varphi^{(2)} = \dfrac{D_{ij}}{6} \dfrac{\partial^2}{\partial X_i \partial X_j  } \dfrac{1}{R_o}
		\label{eq:04c}
	\end{equation}	
	
or, as done for the dipolar term in (\ref{eq:03c}):	
	\begin{equation*}	
		\dfrac{\partial^2}{\partial X_i \partial X_j  } \dfrac{1}{R_o} = \dfrac{X_i X_j}{R_o^5}-\dfrac{\delta_{ij}}{R_o^3}
	\end{equation*}	

	and since $\delta_{ij} D_{ij}=D_{ii} = 0$, we have:
	\begin{equation}	
		\phi^{(2)} = \dfrac{D_{ij} n_i n_j }{2R_o^3}
		\label{eq:04d}
	\end{equation}		
		
	$n_i$ and $n_j$ are the unit vectors starting from $\textbf{R}_o$ and oriented along the two axes of the quadrupolar moment. The eigenvalues of the tensor are such that $D_{ii} = 0$ and are therefore linked by:
	\begin{equation*}	
		D_{xx} = D_{yy}=-\dfrac{1}{2}D_{zz}
	\end{equation*}	
	
	If we write $D$ for component $D_{zz}$, the quadrupolar  potential becomes,  
	\begin{equation}
		\varphi^{(2)} = \dfrac{D}{4\pi R_o^3} (3\textrm{cos}^2 \theta -1) = \dfrac{D}{2R_o^3} \mathcal{P}_2 (\textrm{cos}\ \theta)
		\label{eq:04e}
	\end{equation}
\end{subequations}	
	
	in which one introduces the $\mathcal{P}_2$ \Legendre (\cite{Legendre1785}) polynomial. One can generalize this construction; the \textit{l}-order term is determined by a tensor of order \textit{l}, the 2$^{l}$-polar moment.
	
	The mathematical development above is independent of the physical problem (gravity, magnetism, electricity) one is interested in. It serves to illustrate the fundamental result from \Legendre (\cite{Legendre1785}) regarding the theory of gravitational attraction of masses as generalized by \Laplace (\cite{Laplace1799}): as far as one observes from far enough the sources, the inverse of distance ($\dfrac{1}{\textbf{R}_o - \textbf{r}}$) is given by: 
\begin{subequations}	
	\begin{equation}
	\dfrac{1}{|\textbf{R}_o-\textbf{r}|} = \dfrac{1}{\sqrt{\textbf{R}_o^2 + \textbf{r}^2 - 2\textbf{rR}_o \textrm{cos}\ \chi}} = \sum_{l=0}^{\infty} \dfrac{\textbf{r}^l}{\textbf{R}_o^{l+1}}\mathcal{P}_{l} (\textrm{cos}\ \chi )
		\label{eq:05a}
	\end{equation}
	
	Let us introduce the pairs of spherical angles $\Theta$, $\Phi$  and $\theta$ , $\varphi$  formed respectively by vectors  and  with the given coordinate axes, and apply the addition theorem of spherical functions:
		\begin{equation}
	\mathcal{P}_{l} (\textrm{cos}\ \chi) = \sum _{m=-l}^{l} \dfrac{(l-|m|)!}{(l+|m|)!} \mathcal{P}_{l}^{|m|} (\textrm{cos}\ \Theta) \mathcal{P}_{l}^{|m|} (\textrm{cos}\ \theta) e^{-im(\Phi-\varphi)}
		\label{eq:05b}
	\end{equation}
	
	were $\mathcal{P}_l^m$ are associated Legendre polynomials. Let us also introduce the spherical functions: 
	\begin{equation}
		\mathcal{Y}_{lm} (\theta,\varphi)= (-1)^m i^i \sqrt{\dfrac{(2l+1)(l-m)!}{4\pi(l+m)!}} \mathcal{P}_{l}^{m} (\textrm{cos}\ \theta) e^{im\varphi}, \quad m \ge 0.
		\label{eq:05c}
	\end{equation}
	
	Integrating (\ref{eq:05b}) and (\ref{eq:05c}) in (\ref{eq:05a}), one finally obtains the expression for the inverse of a distance on the sphere: 
		\begin{equation}
			\dfrac{1}{|\textbf{R}_o- \textbf{r}|} = \sum_{l=0}^{\infty} \sum_{m=-1}^{l} \dfrac{r^l}{R_o^{l+1}}\dfrac{4\pi}{2l+1} \mathcal{Y}^{*}_{lm}(\Theta, \Phi) \mathcal{Y}_{lm}(\theta, \varphi) 
		\label{eq:05d}
	\end{equation}
	
	Developing each term in equation (\ref{eq:03a}), one obtains the expression for the term of order \textit{l} of the potential:
	\begin{equation}
			\varphi ^{(l)} = \dfrac{1}{R_o^{l+1}} \sum_{m=-l}^{l} \sqrt{\dfrac{4\pi}{2l+1}}\mathcal{Q}_m^{(l)}\mathcal{Y}^{*}_{(lm)}(\Theta,\Phi)
		\label{eq:05e}
	\end{equation}
	
	where the $2l+1$ quantities $\mathcal{Q}_m^{l}$ constitute the 2$^{l}$-polar moment of the system of charges, defined by:	
	\begin{equation}
		\mathcal{Q}_m^{l} = \sum_i e_i r^l_i \sqrt{\dfrac{4\pi}{2l+1}} \mathcal{Y}_{lm} (\theta_i, \varphi_i)
		\label{eq:05f}
	\end{equation}
\end{subequations}	

	At this point, let us underline why we have taken the trouble to recall this (at least in large part) classical derivation, which is likely taught in all graduate and even undergraduate physics programs. In this section, we have seen that in the case of a static field \textbf{E}, Coulomb’s law (\ref{eq:02e}) imposes itself and the field is radial. In the case of a system of charges, if one is too close to the system the interactions of the charges \textit{forbid one to use the Lagrangian concept of moment}. Thus one must remain \textit{far from the system}. But as found by \Legendre (\cite{Legendre1785}) and \Laplace (\cite{Laplace1799}) when attempting to define the shape of the attraction field of masses, it is seen that the \textbf{E} field involves the same constraints, \ie is \textit{electrostatic}. It is only because we are with a static field that the notion of the \textit{inverse of a distance} takes its full meaning and that we can develop it into \textit{spherical harmonics}.
	
	We can now undertake the same analysis in the case of the magnetic field.
		
	\subsection{The magnetostatic field}\label{sec2-3}
	 As we have seen above, Maxwell’s equations (\ref{eq:01b}) and (\ref{eq:01c}) imply that the magnetic field \textbf{H}, created by charges in finite motion, remaining in a finite region of space (\ref{eq:01b}), whose impulses always retain finite values, has a stationary character that we wish to analyze further.
	 
	The two equations are now:
\begin{subequations}		
	\begin{align}
		\textrm{div}\ \textbf{H} = 0 		\label{eq:06a} \\
		\textrm{rot}\ \textbf{H} = \dfrac{4\pi}{c} \textbf{j} 		\label{eq:06b}
	\end{align}
	
	The vector potential $\textbf{A}$ associated to $\textbf{H}$ is defined by $\textrm{rot}\ \textbf{A} = \textbf{H}$. Carried into (\ref{eq:06b}) it becomes: 
	\begin{equation*}	
	\textrm{grad}\ \textrm{div} \textbf{A} - \Delta  \textbf{A}  = \dfrac{4\pi}{c} \textbf{j} 
	\end{equation*}		
	
	\textbf{A} being defined in a non unequivocal way, one can impose an arbitrary condition, such as $\textrm{div}\ \textbf{A} = 0$. The previous line then becomes the Poisson equation:
	\begin{equation}
		\Delta \textbf{A} = -\dfrac{4\pi}{c} \textbf{j}
		\label{eq:06c}
	\end{equation}

	(\ref{eq:06c}) is analogous to (\ref{eq:02c}), charge density $\rho$ being replaced by current density $\dfrac{\textbf{j}}{c}$. By analogy with the electrical potential, we can write: 
	\begin{equation}
		\textbf{A} = \dfrac{1}{c} \int \dfrac{\textbf{j}}{R} dV	
		\label{eq:06d}
	\end{equation}

	where $R$ is the distance from observation point to volume element $dV$. The integral in (\ref{eq:06d}) can be replaced by a sum and the current density by $\rho \textbf{v}$. By analogy with the scalar potential (\ref{eq:03a}), the vector potential becomes: 
	\begin{equation}
\textbf{A} = \dfrac{1}{c} \sum \dfrac{e_i \textbf{v}_i}{R_i}
		\label{eq:06e}
	\end{equation}
\end{subequations}	
	
	This is how charges are introduced in a vector linked to the magnetic field, without risking the mistake of writing that the magnetic field derives from a scalar potential. 
	
	As done above for the electrostatic field, one can calculate the effect of the moving charges in a reference system with its origin within the charge distribution, with the same notations for vectors $\textbf{r}_i$  and distance $\textbf{R}_o$. (\ref{eq:06e}) becomes:
\begin{subequations}	
	\begin{equation}
		\textbf{A} = \dfrac{1}{c}\sum \dfrac{e_i \textbf{v}_i}{|\textbf{R}_o - \textbf{r}_i|}
		\label{eq:07a}
	\end{equation}

	The main difference between scalar and vector potentials is that for \textbf{E} one only has the effect of fixed charges or motion as a rigid block, whereas for \textbf{H} what counts is the uniform velocity of charges imposed by (\ref{eq:01b}). This is the reason why \Poisson \cite{Poisson1826}’s title is: ‘Du magnétisme en mouvement’ (of magnetism in motion). And this is the main reason why one cannot write a physical description of a magnetic field as a series of spherical harmonics.	
	
	However, some remarkable consequences can be derived. For instance, one can always write in a development as a \Legendre  (\cite{Legendre1785}) series analogous to (\ref{eq:03b}), to first order: 
	\begin{equation*}	
	\textbf{A} = \dfrac{1}{cR_o} \sum e\textbf{v} - \dfrac{1}{c}\sum e\textbf{v}(\textbf{r}\nabla \dfrac{1}{R_o})
	\end{equation*}		
	
	One can write $\sum e {\textbf{v}} = \dfrac{d}{dt} \sum e\textbf{r}$, but the mean value of the derivative varying in a finite interval is 0. 
	\begin{equation}
			\textbf{A} =  - \dfrac{1}{c}\sum e\textbf{v}(\textbf{r}\nabla \dfrac{1}{R_o}) = \dfrac{1}{cR_o^3} \sum e\textbf{v}(\textbf{rR}_o)	
		\label{eq:07b}
	\end{equation}
	
	Note that $\textbf{v} = \dot{\textbf{r}}$, and since $\textbf{R}_o$  is a constant vector,
	\begin{equation*}	
		\sum e(\textbf{R}_o\textbf{r}) \textbf{v}= \dfrac{1}{2}\dfrac{d}{dt} \sum e\textbf{r}(\textbf{R}_o\textbf{r}) + \dfrac{1}{2} \sum e [\textbf{v}(\textbf{r}\textbf{R}_o)  - \textbf{r} (\textbf{v}\textbf{R}_o)]
	\end{equation*}		
	
	Carrying this expression in (\ref{eq:07b}), the mean value of the first term is again 0, thus:
	\begin{equation}
			\textbf{A} = \dfrac{1}{2cR_o^3} \sum e [\textbf{v} (\textbf{rR}_o) - \textbf{r} (\textbf{v} \textbf{R}_o)]	
		\label{eq:07c}
	\end{equation}
	
	One recognizes a vector product in (\ref{eq:07c}). Let us introduce the magnetic moment $\mathfrak{m} = \dfrac{1}{2c} \sum e\textbf{r} \times \textbf{v}$. Equation (\ref{eq:07c}) becomes: 
	\begin{equation}
		\textbf{A} = \nabla \dfrac{1}{R_o} \times \mathfrak{m} = \dfrac{\mathfrak{m}\times \textbf{R}_o}{R_o^3}
		\label{eq:07d}
	\end{equation}

	Whereas $\dfrac{1}{R_o}$ verifies Laplace’s equation, its modification by the rotational of $\mathfrak{m} = \dfrac{1}{2c} \sum e\textbf{r} \times \textbf{v}$ does not. With the expression for the vector potential (\ref{eq:07d}), one can derive the magnetic field. Given the formula $\textrm{rot}\ a \times b = (b\nabla)\ a - (a\nabla)\ b + a\ \textrm{div}\ b - b\ \textrm{div}\ a$, one finds:
	\begin{equation*}	
		\textbf{H} = \textrm{rot}\ \mathfrak{m} \times \dfrac{\textbf{R}_o}{R_o^3} = \mathfrak{m}\ \textrm{div} \dfrac{\textbf{R}_o}{R_o^3} - (\mathfrak{m}\nabla) \dfrac{\textbf{R}_o}{R_o^3}
	\end{equation*}		
	
	Since with $\textbf{R}_o \ne 0$:
	\begin{equation*}	
		\textrm{div} \dfrac{\textbf{R}_o}{R_o^3} = \textbf{R}_o \textrm{grad} \dfrac{1}{R_o^3} + \dfrac{1}{R_o^3} \textrm{div} \textbf{R}_o = 0,
	\end{equation*}
	
	and:
	\begin{equation*}	
		(\mathfrak{m}\nabla) \dfrac{\textbf{R}_o}{R_o^3} = \dfrac{1}{R_o^3}(\mathfrak{m}\nabla) \textbf{R}_o + \textbf{R}_o (\mathfrak{m}\nabla \dfrac{1}{R_o^3} ) = \dfrac{\mathfrak{m}}{R_o^3} - \dfrac{3\textbf{R}_o (\mathfrak{m}\textbf{R}_o)}{R_o^5},
	\end{equation*}
	
	then:
	\begin{equation}	
		\textbf{H} = \dfrac{3\textbf{n} (\mathfrak{m}\textbf{n})-\mathfrak{m}}{R_o^3},
		\label{eq:07e}
	\end{equation}
\end{subequations}	
	where \textbf{n} is the unit vector along direction $\textbf{R}_o$. If the ratio of mass to charge is the same for all charges in the system, then: 
		\begin{equation*}
			\mathfrak{m} = \dfrac{1}{2c} \sum e\textbf{r}\times\textbf{v} = \dfrac{e}{2mc} \sum m\textbf{r} \times \textbf{v} 
		\end{equation*}
	
	Finally, if all velocities of all charges are such that $v \ll c$, then $m\textbf{v}$ is the impulsion \textbf{p} of the charge and: 
\begin{subequations}		
		\begin{equation}	
			\mathfrak{m} = \dfrac{e}{2mc} \sum \textbf{r} \times \textbf{p} = \dfrac{e}{2mc}\mathcal{M}
		\label{eq:08a}
	\end{equation}
	
	where $\mathcal{M} = \sum \textbf{r} \times\textbf{p}$ is the kinetic moment of the system. Here the ratio of magnetic to mechanical moment is a constant.
	
	Let us now consider a system of charges placed in a constant, external magnetic field \textbf{H}. The (time averaged) force exerted on the system is the Lorentz force. Following \blue{Lorentz} (\cite{Lorentz1898}), it is known as \blue{Maxwell}’s 5ft equation and given by $\mathcal{F} = \sum \dfrac{e}{c} \textbf{v}\times\textbf{H} = \dfrac{d}{dt} \sum \dfrac{e}{c} \textbf{r} \times \textbf{H}$, that is zero. On the other hand, the time average of the moment of forces is $\mathcal{K} = \sum \dfrac{e}{c} \textbf{r}\times(\textbf{v}\times \textbf{H})$\textit{ which is not 0}. Writing explicitly the double vector product,
	\begin{equation*}	
		\mathcal{K} = \sum \dfrac{e}{c} \{\textbf{v}(\textbf{rH}) - \textbf{H}(\textbf{vr})\} = \sum\dfrac{e}{c}\{\textbf{v}(\textbf{rH}) -\dfrac{1}{2} \textbf{H} \dfrac{d}{dt} \textbf{r}^2\}
	\end{equation*}
	 
	one obtains simply:
	\begin{equation*}	
		\mathcal{K} = \mathfrak{m} \times \textbf{H}	 
	\end{equation*}
	
	 The Lagrangian of a system of charges placed in an external, constant and uniform magnetic field is:  
	\begin{equation}
		\mathcal{L}_{H} = \sum \dfrac{e}{c} \textbf{Av} = \sum \dfrac{e}{2c} (\textbf{H}\times \textbf{r})\textbf{v} = \sum \dfrac{e}{2c} (\textbf{r}\times \textbf{v}) \textbf{H}	 
		\label{eq:08b}
	\end{equation}		
	
	Introducing the magnetic moment (\ref{eq:08b}) becomes simply,
	\begin{equation}
		\mathcal{L}_{H} = \mathfrak{m} \textbf{H}	 
	 	\label{eq:08c}
	\end{equation}		
	By analogy, in a uniform electric field, the Lagrangian of a system with 0 total charge and a dipolar moment includes the term: 
	\begin{equation}
		\mathcal{L}_{E} = \textbf{dE}
	 	\label{eq:08d}
	\end{equation} 
	
	Let us consider a system of charges undergoing a finite motion in a field \textbf{E} with central symmetry, due to a motionless particle. Let us shift from a motionless system of coordinates to a system undergoing a uniform rotation about an axis passing through the motionless particle. The velocity \textbf{v} of the particle in the new system is linked to its velocity \textbf{v} in the old system by $\textbf{v}^{'} = \textbf{v} + \Omega \times \textbf{r}$ where $\textbf{r}$ is the particle’s vector radius and $\Omega$ the angular velocity of the rotating coordinate system. In the fixed system, the Lagrangian of charges is:
	\begin{equation*}	
		\mathcal{L} = \sum \dfrac{mv^{'2}}{2}-\mathcal{U},	 
	\end{equation*}

	where $\mathcal{U}$ is the potential energy of charges in the external field E. In the new system the Lagrangian becomes: 	 
	\begin{equation*}	
		\mathcal{L} = \sum \dfrac{m}{2} (\textbf{v}+\Omega \times \textbf{r})^2 - \mathcal{U}.
	\end{equation*}
	
	If the ratio $\dfrac{e}{m}$ of charge to mass is the same for all particles, and if one writes:
 	\begin{equation*}	
		\Omega = \dfrac{e}{2mc}\textbf{H}	 
	\end{equation*}	
	
	then, for sufficiently small values of \textbf{H} so that one can neglect terms in $\textbf{H}^2$, the Lagrangian takes the form:
	\begin{equation}
		\mathcal{L} = \sum \dfrac{mv^2}{2} + \dfrac{1}{2c}\sum e (\textbf{H} \times \textbf{r})\textbf{v} - \mathcal{U}
		\label{eq:08e}
	\end{equation}	 

	 It is remarkable that the two Lagrangians in (\ref{eq:08b}) and (\ref{eq:08c}) are the same. In summary, the Lagrangian of charges in finite motion in an electric field produced by a motionless rotating particle (\ref{eq:08c}) is the same as the Lagrangian of a system of charges placed in a constant and uniform magnetic field (\ref{eq:08b}). In slightly different, more readable terms, the behavior of a system of charges with the same $\dfrac{e}{m}$ ratio executing a finite motion in a field \textbf{E} with central symmetry and a weak and uniform field \textbf{H} is equivalent to the behavior of the same charge system in field \textbf{E} with respect to a uniformly rotating coordinate system with angular velocity $\Omega$.  This is \blue{Larmor} ’s theorem (\cf \cite{Larmor1900}, chapter VI, \cite{Brillouin1945}). 	 
	 
	 Let us now evaluate the variation of the mean kinetic moment $\mathcal{M}$. The variation of the mean kinetic moment $\mathcal{M}$ is equal to the moment $\mathcal{K}$ of forces applied to the system. Thus $\mathcal{K} = \mathfrak{m} \times \textbf{H} = \dfrac{d\mathcal{M}}{dt}$.  If the ratio $\dfrac{e}{m}$ is the same for all the system’s particles, then $\mathcal{M}$ is proportional to $\mathfrak{m}$ (\cf \ref{eq:08a}), and: 
	\begin{equation}
		\dfrac{d\mathcal{M}}{dt} = -\Omega \times \mathcal{M}.
		\label{eq:08f}
	\end{equation}	
	 
	 Vector $\mathcal{M}$, and thus vector $\mathfrak{m}$, both rotate with angular velocity $-\Omega$ about the field direction; its absolute value and the angle it makes with respect to the field direction are constant.
\end{subequations}	

\section{Some further remarks on section \ref{sec2}}\label{sec3}
	Most if not all modern studies of the geomagnetic field have it varying in time, keeping its first order dipolar configuration and ‘ready’ to be analyzed with spherical harmonics. But a time variable \textbf{E} or \textbf{H} field implies wave propagation, hence physics described by Helmholtz equations, not Legendre. Let us describe what happens with a variable field in the equations of section \ref{sec2}. 
	
	The only link between the field geometry and the physics of the problem, that is the e charges and their positions in the reference system, is (\ref{eq:05f}) that we recall here:
	\begin{equation*}	
		\mathcal{Q}_m^{l} = \sum_i e_i r^l_i \sqrt{\dfrac{4\pi}{2l+1}} \mathcal{Y}_{lm} (\theta_i, \varphi_i)
	\end{equation*}	
	
	\begin{figure}[H]
		\centering{\includegraphics[width=1\columnwidth]{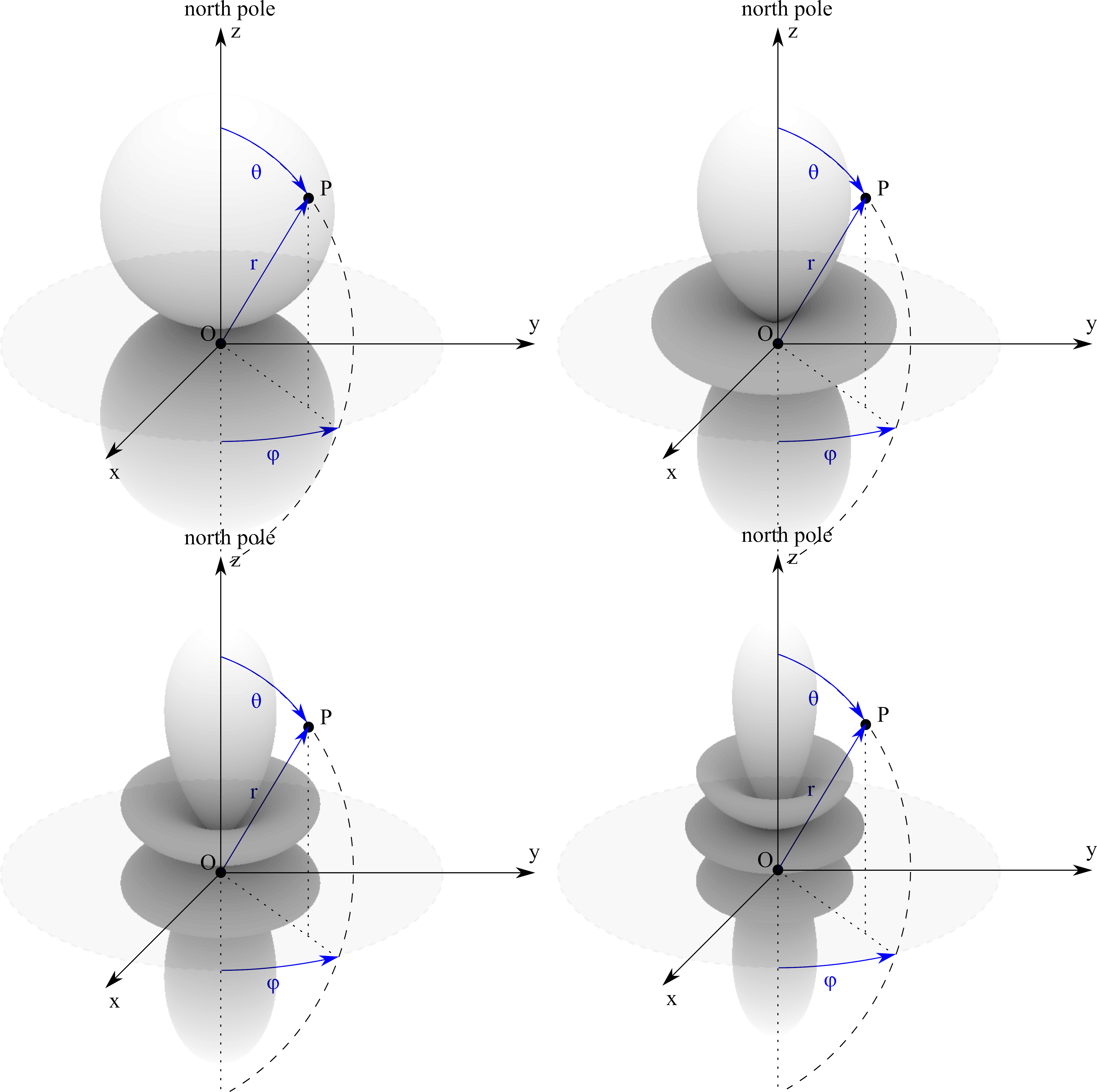}} 	
     	\caption{Spherical harmonics (from left to right and top to bottom) $\mathcal{Y}_{1,0}$, $\mathcal{Y}_{2,0}$, $\mathcal{Y}_{3,0}$ and $\mathcal{Y}_{4,0}$.}
		\label{fig:01} 	
	\end{figure}	  

	One needs to visualize the shapes of spherical harmonics, that are “physically” represented by the coefficients $\mathcal{Y}_{lm}$ . Figure \ref{fig:01} displays the first four axial multipoles, that is the eigenvectors $\mathcal{Y}_{1,0}$, $\mathcal{Y}_{2,0}$, $\mathcal{Y}_{3,0}$ and $\mathcal{Y}_{4,0}$ (from left to right and top to bottom) corresponding to the so-called Gauss coefficients $g_{1,0}$, $g_{2,0}$, $g_{3,0}$ and $g_{4,0}$. The eigenvectors are a constant of the problem, as long as this representation retains a physical meaning. For example, a dipolar magnetic field must always have its principal axis aligned with the axis going from the origin to the geographic North pole. In this paradigm, a variable field would involve only modifications of the space-time physics of term $e^{i}r^{l}_{i}$ in the sum (\ref{eq:05f}). But this constraint cannot be satisfied because either:
\begin{enumerate}
	\item the position term $r_i^{l}$ of each moving particle fluctuates with time, so that the Legendre-Laplace condition (05a) is not satisfied any more, that is the inverse distance $\dfrac{1}{|\textbf{R}_o - \textbf{r}|}$ is no more a natural solution of the Laplacian. One would need to introduce time, but then the Laplacian would have to be replaced by a Dalembertian, i.e. a different problem;
	
	\item or it is the number and/or quality of the charges that would change with time. But then the nature of the core would change with time and one would need to find a physical mechanism that would explain how the field intensity could decrease (as is the case at present), yet could have increased and even reversed in the past.
\end{enumerate}

	\Poisson \cite{Poisson1826} reasoning was simpler. As early as page 7 of his memoir, above his first equation, he linked gravitation and magnetism. So did \Gauss (\cite{Gauss1837}), and this was much developed by \blue{Heaviside} (\cite{Heaviside1893}). This was more than a simple physical analogy.
	
	Recall that modern geophysicists use \blue{Talwani}'algorithms (\cite{Talwani1959,Talwani1965}) to determine the mass (respectively dipolar) distributions based on microgravimetric (resp. magnetic) measurements using the same equation. \Poisson (\cite{Poisson1826}) writes: '\textit{The differential volume element, corresponding to point M’ [The magnetic volume], will have $h^3 d\chi d\xi d\eta$ as its expression; we will write $\mu^{‘} h^3 d\chi d\xi d\eta$ for the amount of free fluid it contains, $\mu^{'}$ being positive or negative according to this fluid being boreal or austral. This coefficient will be a function of $\chi, \xi, \eta$ depending on the distribution of the two fluids inside the magnetic elements. If they are moving, $\mu^{'}$ will vary with time; but the total quantity of free fluid belonging to the same element must always be zero, so we will always have}: 
	\begin{equation*}	
			\int \mu^{‘} d\chi d\xi d\eta = 0 \qquad (I)
	\end{equation*}	
\textit{	The integral extending to the entire volume of the magnetic element'.}
	
	This short quotation is from the very beginning of \Poisson (\cite{Poisson1826}) derivation; in more modern terms \textrm{div} \textbf{H} = 0. What enters a volume element and what leaves it is constant: the field is stationary.
	
	From \Lagrange 's standpoint (\cite{Lagrange1788}), \Poisson (\cite{Poisson1826}) implies that at the origin of the magnetic field is a rather undeformable object, a solid body consisting of a set of electromagnetic point sources whose respective distances do not vary, or that can be considered as such at the distance at which its effects are observed. We note that the tensor of a quadrupolar electrical moment $D_{ij} = \sum e (3x_i x_j – r^2 \delta_{ij})$ is similar to the order 2 inertia moment tensor of a rotating solid body $I_{ik} = \sum m (x_{i}^{2}\delta_{ik} – x_{i} x_{k})$.
	
	One then concludes that the space-time variations of electric charges that would not result in a constant field would make \textit{it impossible to construct either an electric or a magnetic moment}.
	
\section{Reconciling modern observations with Poisson’s theory}\label{sec4}	
	\subsection{On the drift of the magnetic dipole}\label{sec4-1}
	In sections \ref{sec2} and \ref{sec3}, we have shown that the only way a magnetic field can be written as the sum of multipolar potentials is that (similar to masses composing a solid rotating body) charges generating the electric and magnetic fields should move uniformly in space and time. In other words the field must be stationary. We also know that the field intensity fluctuates and that this secular variation is morphologically similar to the drift of the rotation pole, called the Markowitz (or Markowitz-Stoyko, \cite{Markowitz1968,Stoyko1968}) drift. The annual oscillations of the field are morphologically similar to those of the length of the day (\cite{Jault1988,Jault1989,Jault1990,Jault1991}).	
	
\begin{figure}[H]
	\centering{\includegraphics[width=1.1\columnwidth]{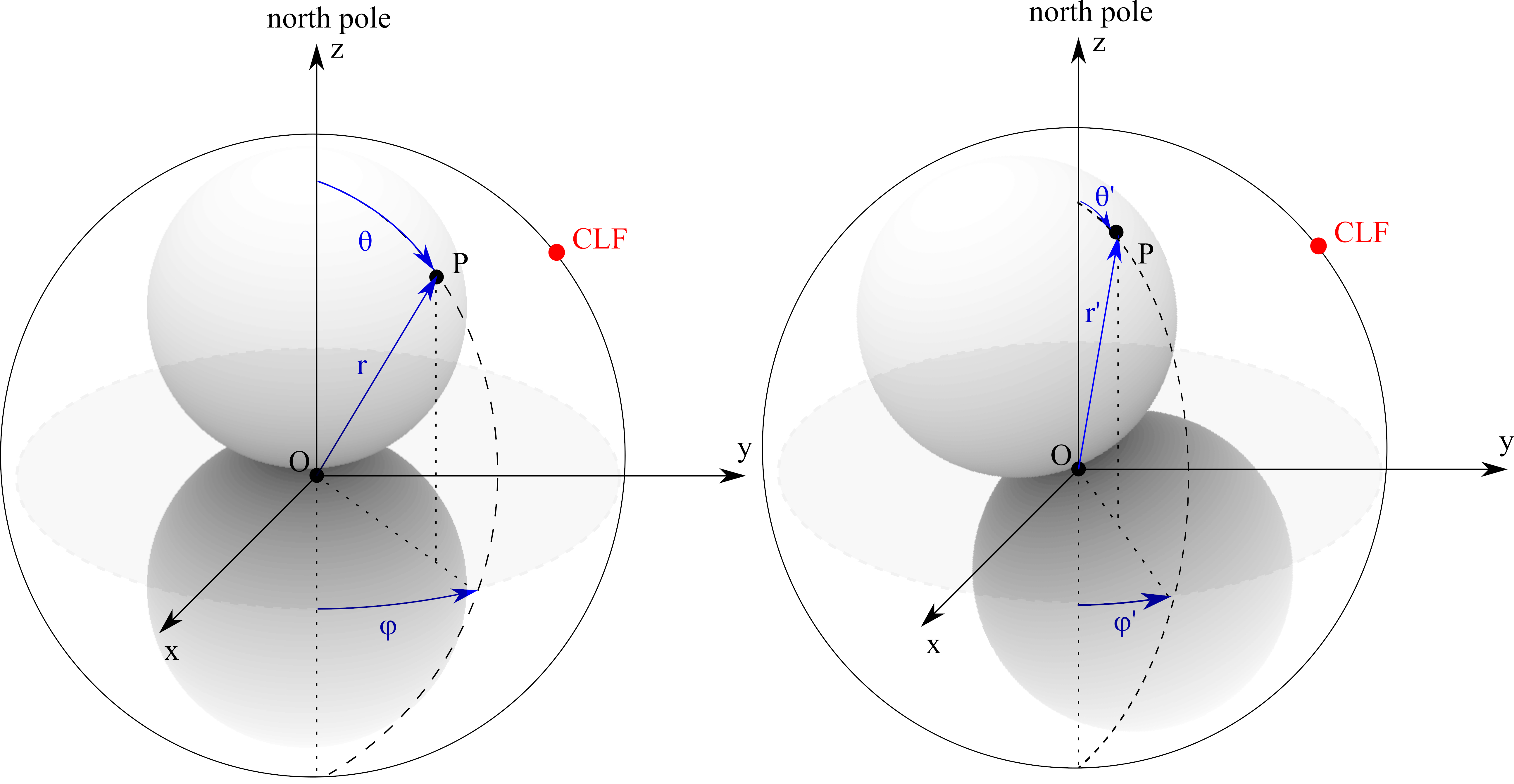}} 	
    \caption{Eigen vector $\mathcal{Y}_{1,0}$ associated with Gauss coefficient $g_{1,0}$: to the left the dipole is axial}
	\label{fig:02} 	
\end{figure}

	One can reconcile \Poisson (\cite{Poisson1826})  and observations by picturing a dipole that oscillates about geographical North, this oscillation sharing the same excitation (forcings) that act on polar motion. Let us first illustrate this idea. In Figure \ref{fig:02} we show the eigenvector $\mathcal{Y}_{1,0}$ associated with the Gauss coefficient $g_{1,0}$. \textbf{CLF} stands for the Chambon-La-Forêt magnetic observatory. Point P is one of the elements of \cite{Poisson1826}'s magnetic volume composing the dipole. As seen in section \ref{sec2}, the dipole action at \textbf{CLF} is characterized by the distance \textbf{P-CLF} on the sphere.  If the dipole is tilted (Figure \ref{fig:02}, right) this distance changes, and so do the coordinates of \textbf{CLF} in the two dipole reference systems (Figure \ref{fig:02}, left and right). 
	
	Following \Lagrange (\cite{Lagrange1788}), a number of authors (\eg \cite{Milankovic1920,Laskar2011,Lopes2021,Bank2022,Lopes2022b,Lopes2023}) have shown how and how much astronomical forces influence Earth’s rotation, in the same way its own weight perturbs the rotation of a spinning top, through an exchange of angular moments. The same astronomical forces perturb the number of sunspots, \ie solar activity (\eg \cite{Morth1979,Scafetta2012,Stefani2016,Courtillot2021}). Since \cite{Laplace1799}, we know that masses at the surfaces of planets can re-organize under the influence of stresses acting on the rotation pole. The \blue{Liouville-Euler} system of linear differential equations of first order runs this re-organization (\cf \cite{Lambeck2005}, section 3 for more details). From sections \ref{sec2} and \ref{sec3}, moving charges may be considered as a moving fluid in rotation about the dipole’s symmetry axis.Any large scale fluid motions must correspond to a block rotation about the Earth’s rotation axis, as shown by Laplace and Poincaré. There are no other possibilities of natural motion at first order. Inside as well as at its surface the pattern of fluid motion must be the same (\cf \cite{Laplace1799,Poincare1885}; see \cite{Courtillot2022,Lopes2022a}, for some illustrations).	
	
	Possibly the longest series of magnetic measurements is that of declination D compiled for Paris (France) by \blue{Alexandrescu et \textit{al.}} (\cite{Alexandrescu1996,Alexandrescu1997}). The series starts in 1541 but we select values from 1781 onward, when the sampling becomes more regular. The Brest sea level data start in 1807. The polar motion data are defined as the modulus of the horizontal displacement of the pole in a plane tangential to the Earth ($m_{1}+i\ m_{2}$); it starts in 1846 and therefore sets the length of our analyzes.
	
	We have applied Singular Spectrum Analysis (\textbf{SSA}, \eg \cite{Golyandina2013}) in order to extract the trend (this sub-section) and the annual and semi-annual components (next subsection) from the three time series of sea-level at Brest, magnetic field at \textbf{CLF} and length of day. \textbf{SSA} uses the mathematical properties of descending order diagonal matrices (Hankel, Toeplitz matrices, \eg \cite{Lemmerling2001}) and their orthogonalization by singular value decomposition (\textbf{SVD}, \eg \cite{Golub1971}).
	
\begin{figure}[H]
     \centering
     \begin{subfigure}[b]{0.5\textwidth}
         \centering
         \includegraphics[width=1\columnwidth]{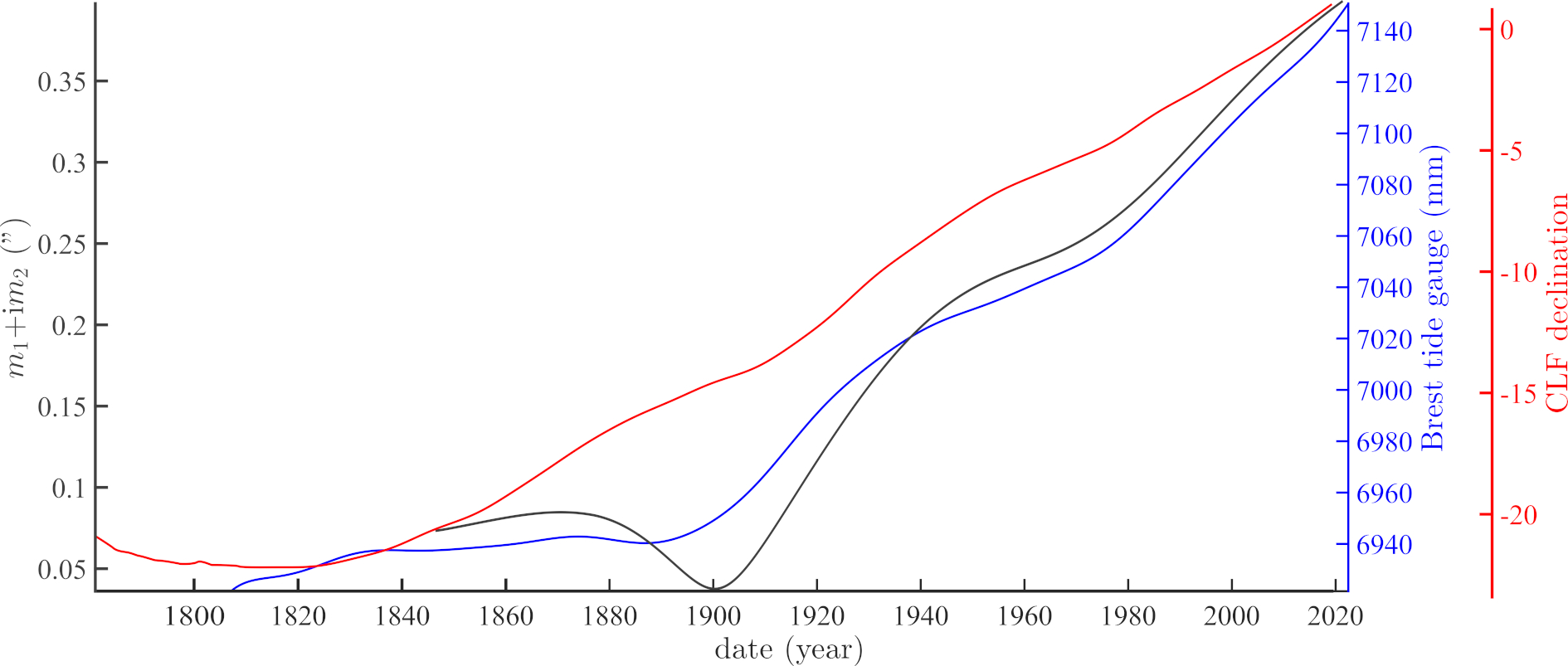}	
         \caption{Superposition of \textbf{SSA} trends of 1) the Markowitz-Stoyko drift since 1846 (gray curve) 2) of the magnetic declination $\mathcal{D}$ in Paris since 1781 (red curve) and 3) of the mean sea level from Brest tide gauge since 1807 (blue curve).}
         \label{fig:03a}
     \end{subfigure}
     \vfill
	 \ \\
     \begin{subfigure}[b]{0.5\textwidth}
         \centering
         \includegraphics[width=1\columnwidth]{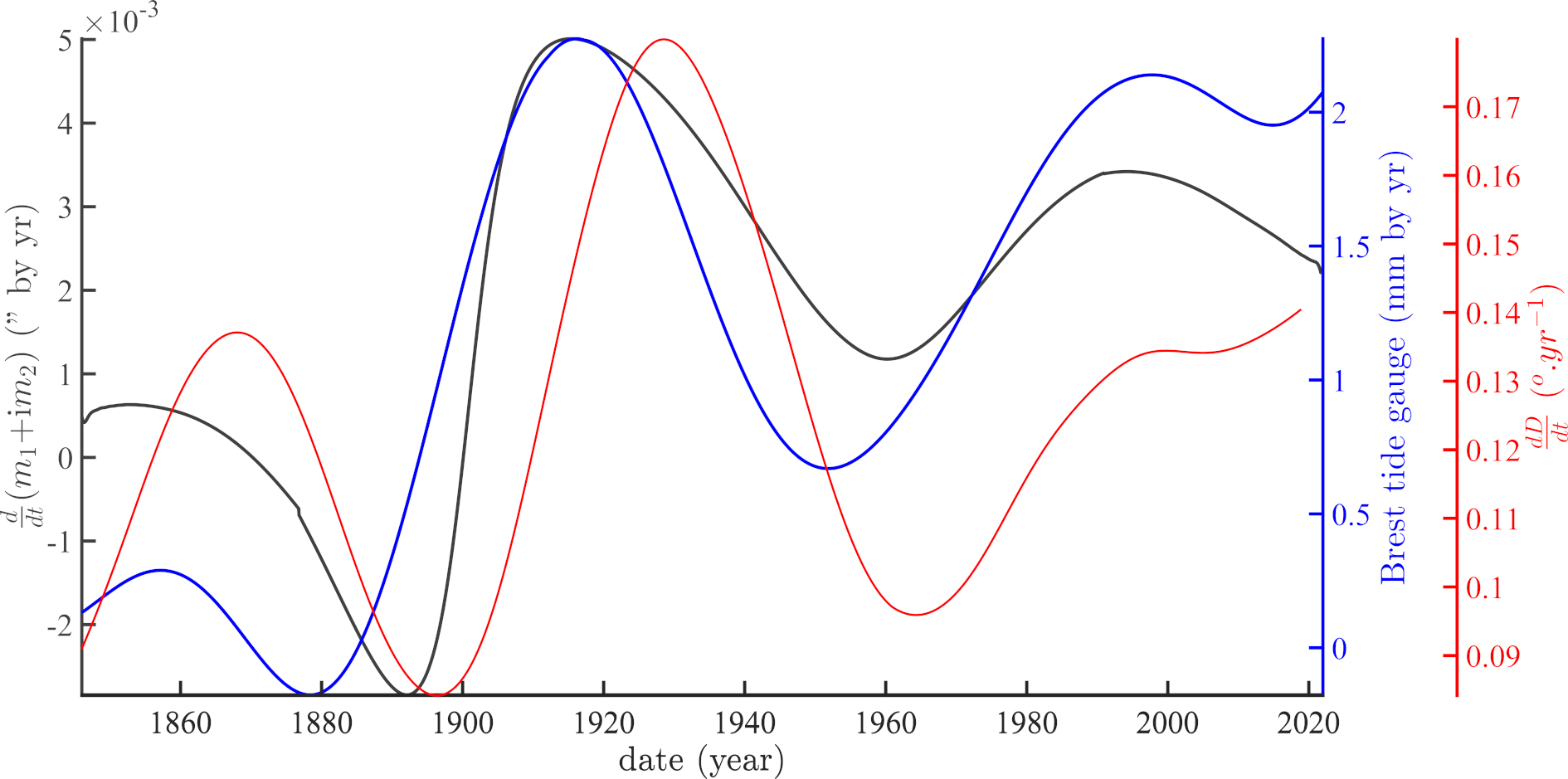} 	
         \caption{Superposition of the first time derivatives of the three trends in Figure \ref{fig:03a}}
         \label{fig:03b}
     \end{subfigure}
          \vfill
		  \ \\
     \begin{subfigure}[b]{0.5\textwidth}
         \centering
         \includegraphics[width=1\columnwidth]{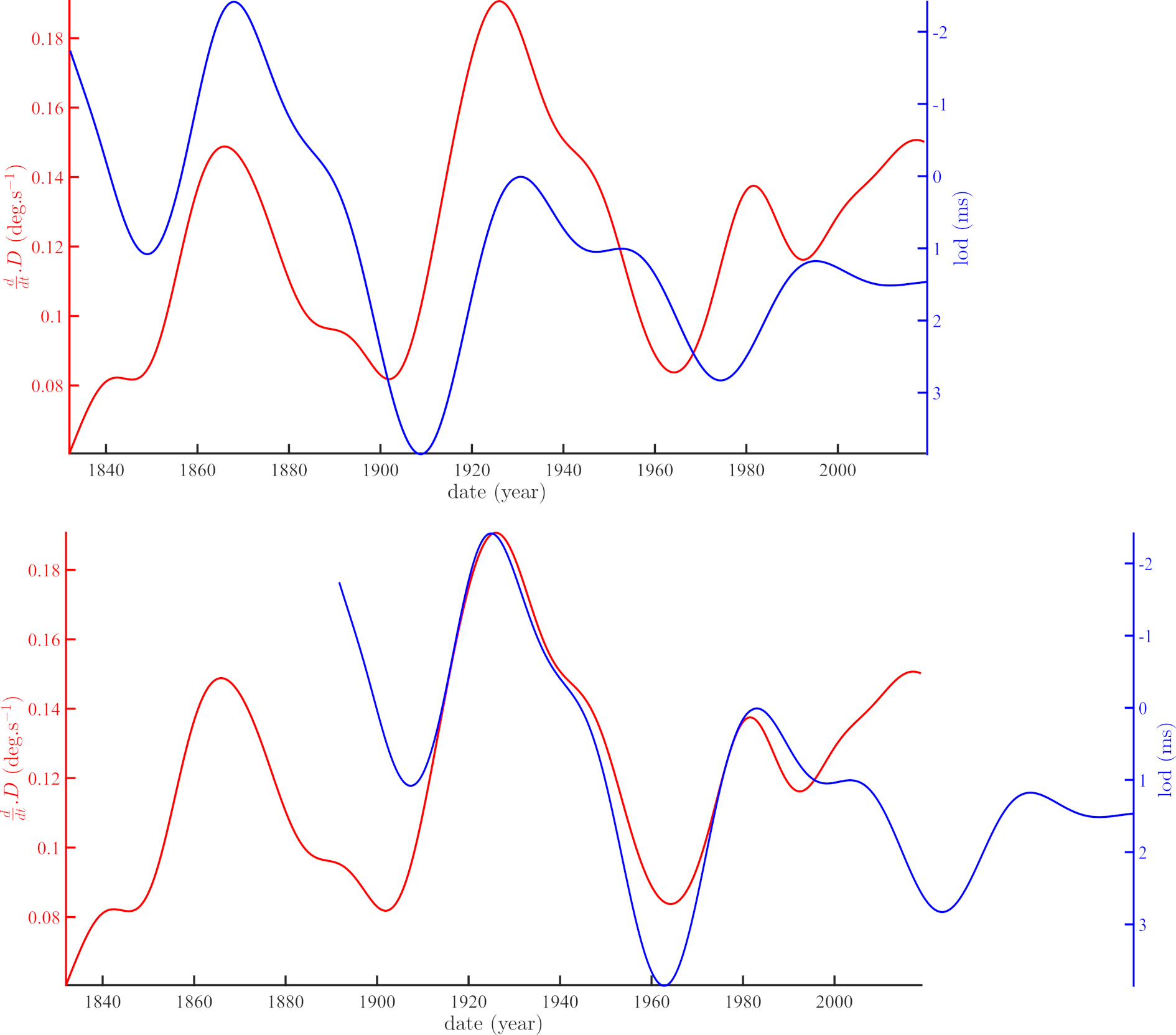} 	
         \caption{On the top: superposition of the smoothed first time derivative of magnetic declination $\mathcal{D}$ in Paris since 1835 (red curve) and the smoothed length of day (blue curve) from \blue{Stephenson and Garrison} (\cite{Stephenson1984}) since 1835. On the bottom: the latter curve has been offset by 60 yr.}
         \label{fig:03c}
     \end{subfigure}
     \caption{Comparison between the trends variations of the magnetic declination in Paris, of the mean sea level recorded in Brest, of the polar motion drift and of the length of day.}
     \label{fig:03}
\end{figure}

	In Figure \ref{fig:03a}, we superimpose the \textbf{SSA} trends of declination in \textbf{CLF} in red, sea level at Brest in blue and polar motion in black. We have dealt with the sea-level series in \blue{Le Mouël et \textit{al.}} (\cite{LeMouel2021}), in which we compared variations of the sea-level trends with the variations of \blue{Markowitz-Stoyko} polar drift. We have studied many aspects of polar motion in \cite{Lopes2017,Lopes2021,Lopes2022c}. The time derivatives of the three trends are shown in Figure \ref{fig:03b}. We have also used the long data set of \blue{Stepheson and Morrison} (\cite{Stephenson1984}) and \blue{Gross} (\cite{Gross2001}) for \textit{lod} and compare its trend with the first derivative of magnetic declination (Figure \ref{fig:03c}). 	
	
	The trends of the three derivatives display very similar patterns. We have shown before that this pattern is linked to the ephemerids of Uranus (\cite{LeMouel2021,Lopes2021}). The similarity between the derivative of the trend of the rotation pole and sea-level is expected: as the Earth shifts, its fluid envelope shifts as a solid in the same way. The two are almost in phase. For the magnetic field, there is a $\sim$20-year phase lag. It is in quadrature with polar motion around 1930, and catches up in the 1960s. This vindicates the mechanism advocated by \blue{Le Mouël et \textit{al.}} (\cite{LeMouel1984,Jault1988}) and \blue{Jault et \textit{al.}} (\cite{Jault1989,Jault1990,Jault1991}), that is a transfer of moment at the core-mantle boundary. The phase lag would be due to the roughness of the \textbf{CMB}.
	
	\Laplace (\cite{Laplace1799}) has shown theoretically that lod and pole motion are linked by a first order derivative operator. We have verified this with observations in \blue{Lopes et al.} (\cite{Lopes2022c}). Figure \ref{fig:03c} compares the (5 yr smoothed) mean value of the derivative of declination in Paris to the corresponding (5 yr smoothed) mean value of length of day. The two series of mean values of solid Earth motions are in quadrature and show a ~60 yr oscillation (\cite{Lopes2022c}, Figure 04). This 60 yr period is clearly present in the 2-humps pattern of lod as well as in the derivative of declination (blue curve, Figure \ref{fig:03c}). As shown in section 2, the mechanical and magnetic moments that are at the origin of the geomagnetic field are linked linearly (equation \ref{eq:08c}). On another hand, the mechanical moment and the polar rotation axis are linked through a first order time derivative (equation \ref{eq:08f}). In the spirit of Larmor’s theorem, we assume that the two phenomena (rotation and magnetism) are linked by the same external forcing ; thus the former should be compared to the derivative of the latter. In the lower part of Figure \ref{fig:03c}, the mean value of lod has been offset by 60 years, leading to an almost perfect match of the patterns. We have not (yet ?) been able to explain that offset.
\newpage
	
	One cannot envision that the magnetic field (declination) variations would be due to intensity variations of components X, Y and Z since as seen in section \ref{sec2} the field must be constant (following \Poisson). Figure \ref{fig:03b} is but an extension to sea-level of an observation made by \LeMouel \cite{LeMouel1984}.
		
	\subsection{On the forced quasi-cycles of the magnetic field}\label{sec4-2}
\begin{figure}[H]
     \centering
     \begin{subfigure}[b]{0.5\textwidth}
         \centering
         \includegraphics[width=1\columnwidth]{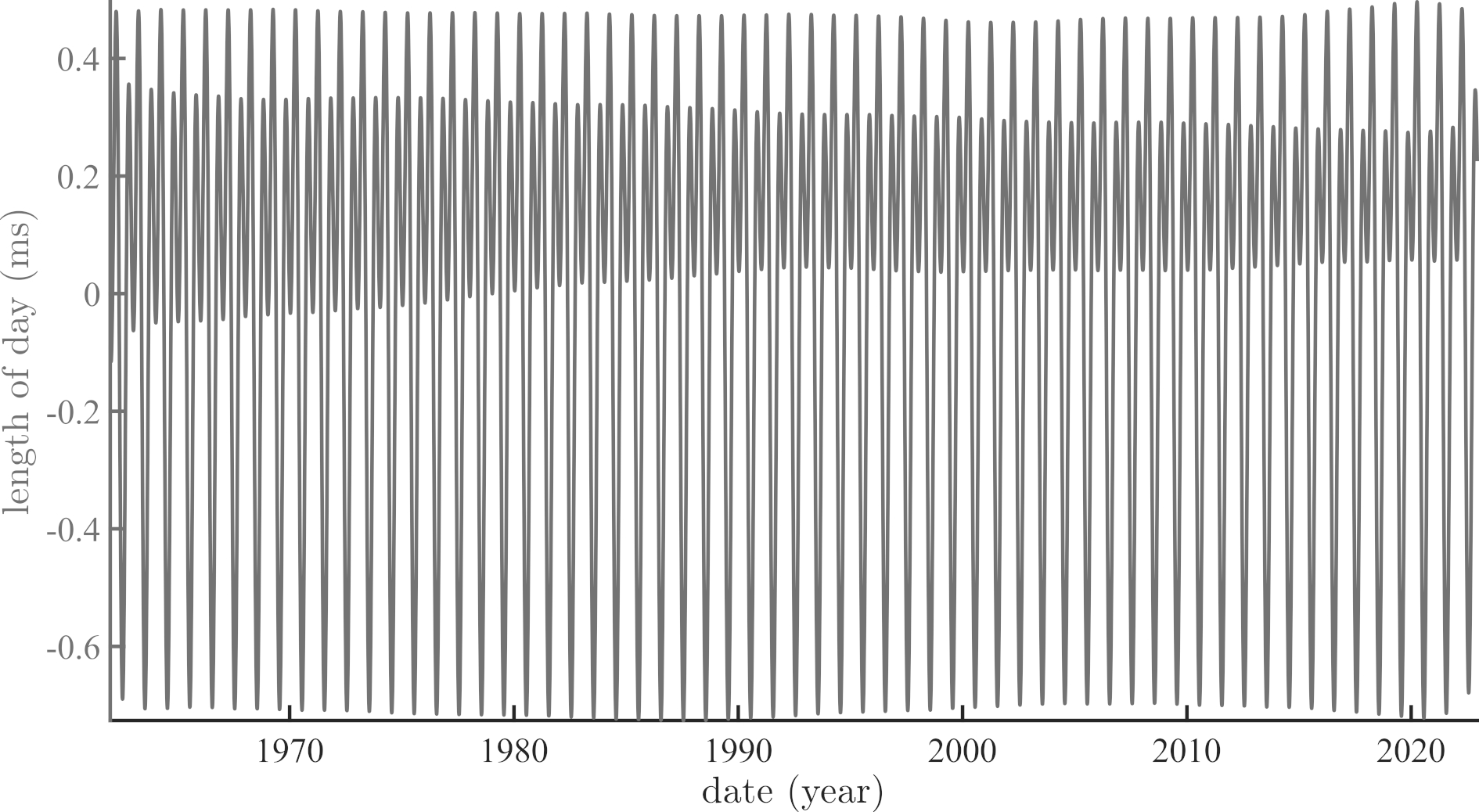}	
         \caption{Sum of the annual and semi-annual components of sea-level at the Brest tide gauge since 1962, extracted by \textbf{SSA}.}
         \label{fig:04a}
     \end{subfigure}
     \vfill
     \begin{subfigure}[b]{0.5\textwidth}
         \centering
         \includegraphics[width=1\columnwidth]{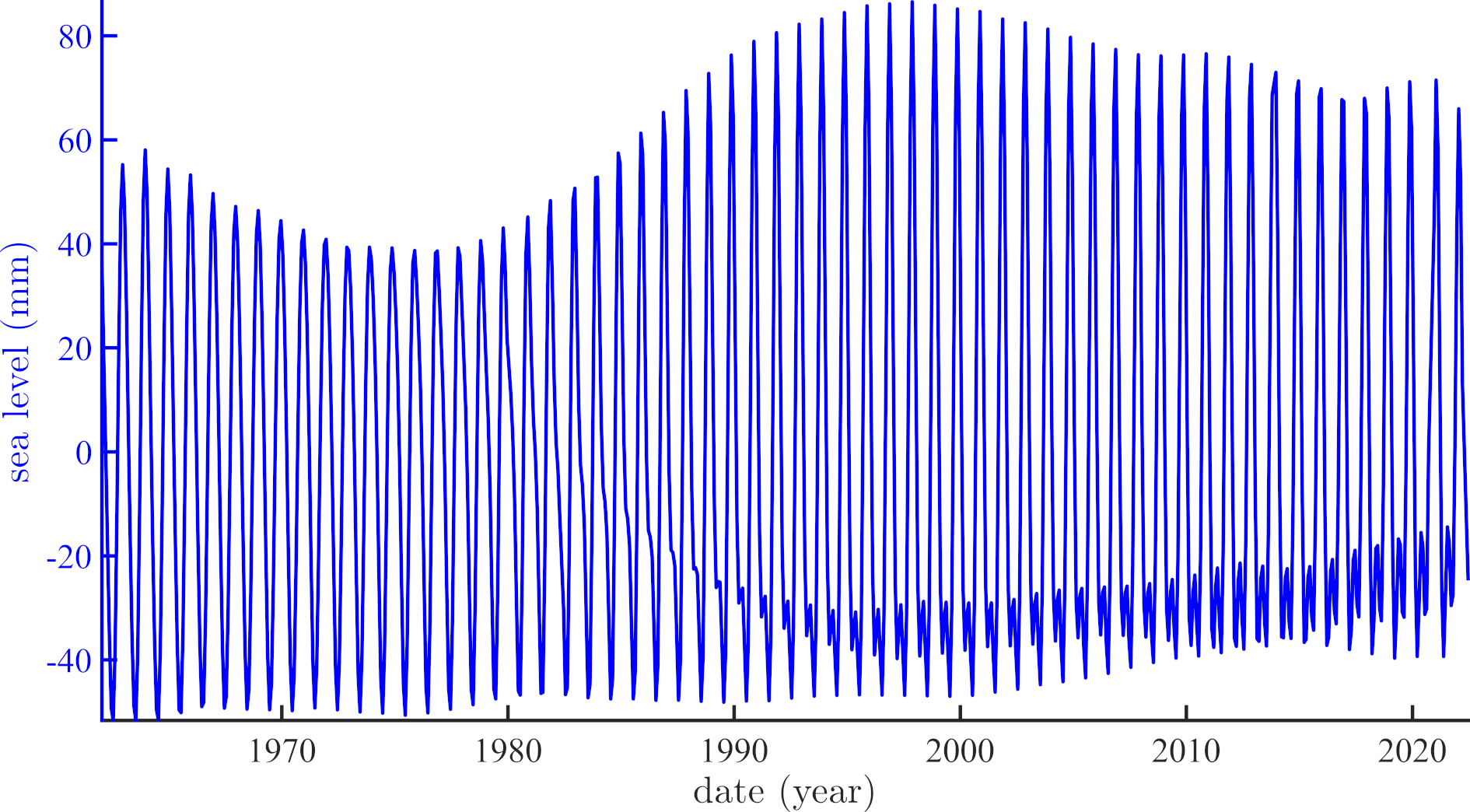} 	
         \caption{Sum of the annual and semi-annual components of the X magnetic component at CLF since 1962, extracted by \textbf{SSA}.}
         \label{fig:04b}
     \end{subfigure}
     \vfill
     \begin{subfigure}[b]{0.5\textwidth}
         \centering
         \includegraphics[width=1\columnwidth]{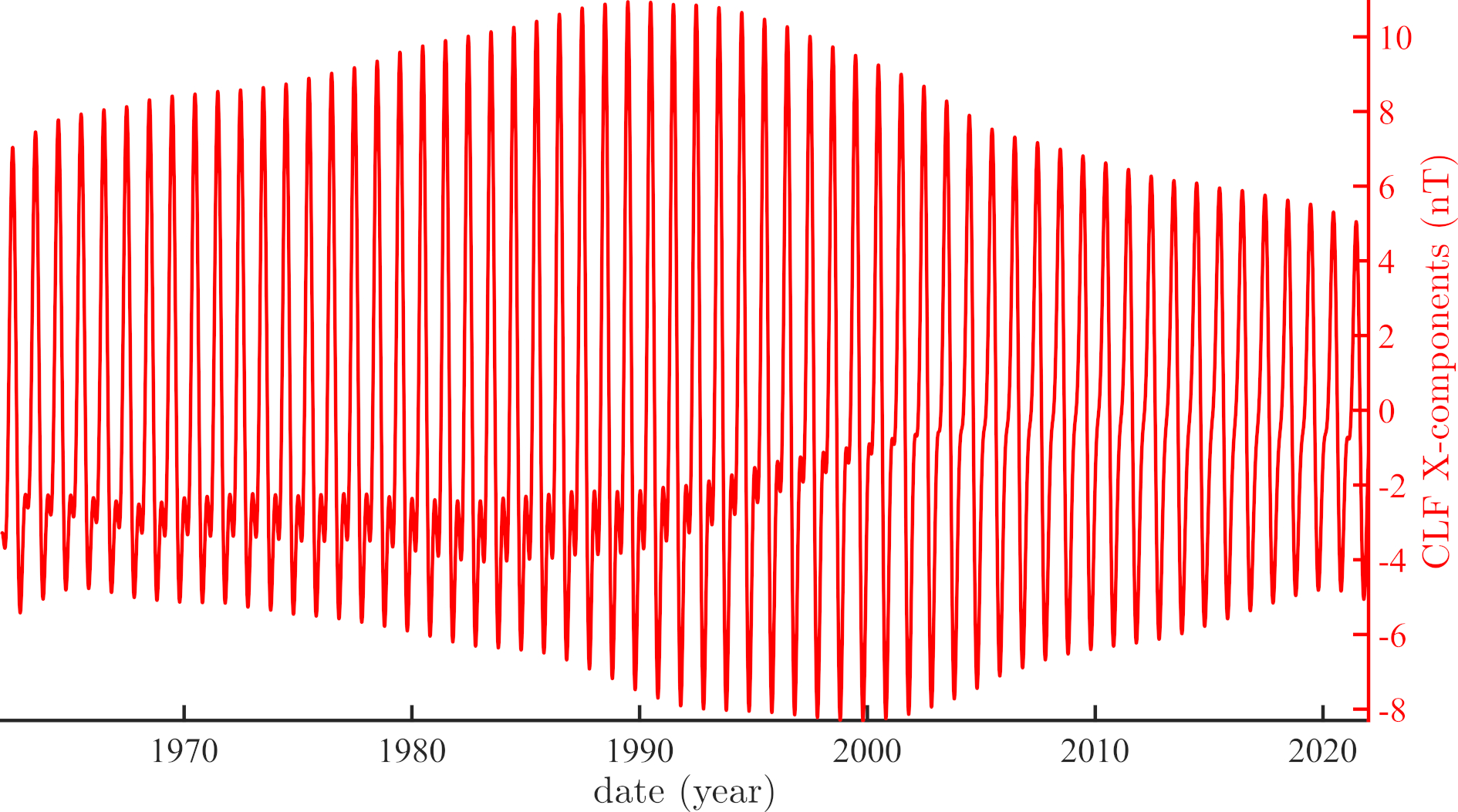} 	
         \caption{Sum of the annual and semi-annual components of lod since 1962, extracted by \textbf{SSA}.}
         \label{fig:04c}
     \end{subfigure}
     
     \caption{Trends (a) and (b) first time derivatives of trends of polar motion drift (black curve), magnetic declination at CLF (red curve) and sea level at Brest (blue curve).}
     \label{fig:04}
\end{figure}

	Next, we extend the observation by \blue{Jault et Le Mouël} (\cite{Jault1991}) of a link between annual and semi-annual oscillations of lod and the magnetic field. We have applied \textbf{SSA} in order to extract the annual and semi-annual components from the same three time series (sea-level at Brest, magnetic field at \textbf{CLF} and length of day; this sub-section and Figures \ref{fig:04a}, \ref{fig:04b}, \ref{fig:04c}). 	
	
\begin{figure}[H]
     \centering
     \begin{subfigure}[b]{0.5\textwidth}
         \centering
         \includegraphics[width=1\columnwidth]{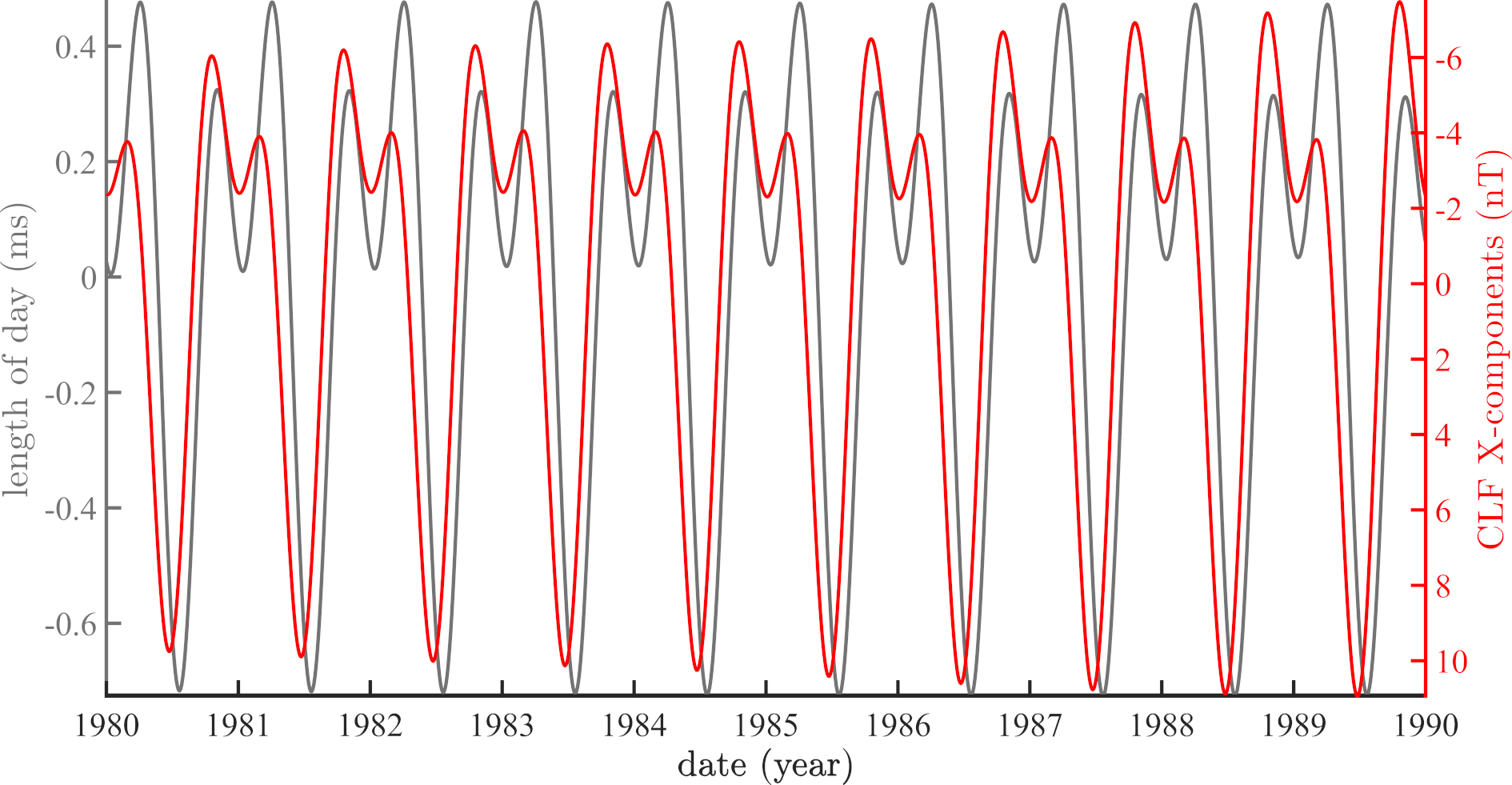}	
         \caption{ Enlargement and superposition of the annual and semi-annual components of the X magnetic component at \textbf{CLF} (red) and length of day (gray).}
         \label{fig:05a}
     \end{subfigure}
     \vfill
     \begin{subfigure}[b]{0.5\textwidth}
         \centering
         \includegraphics[width=1\columnwidth]{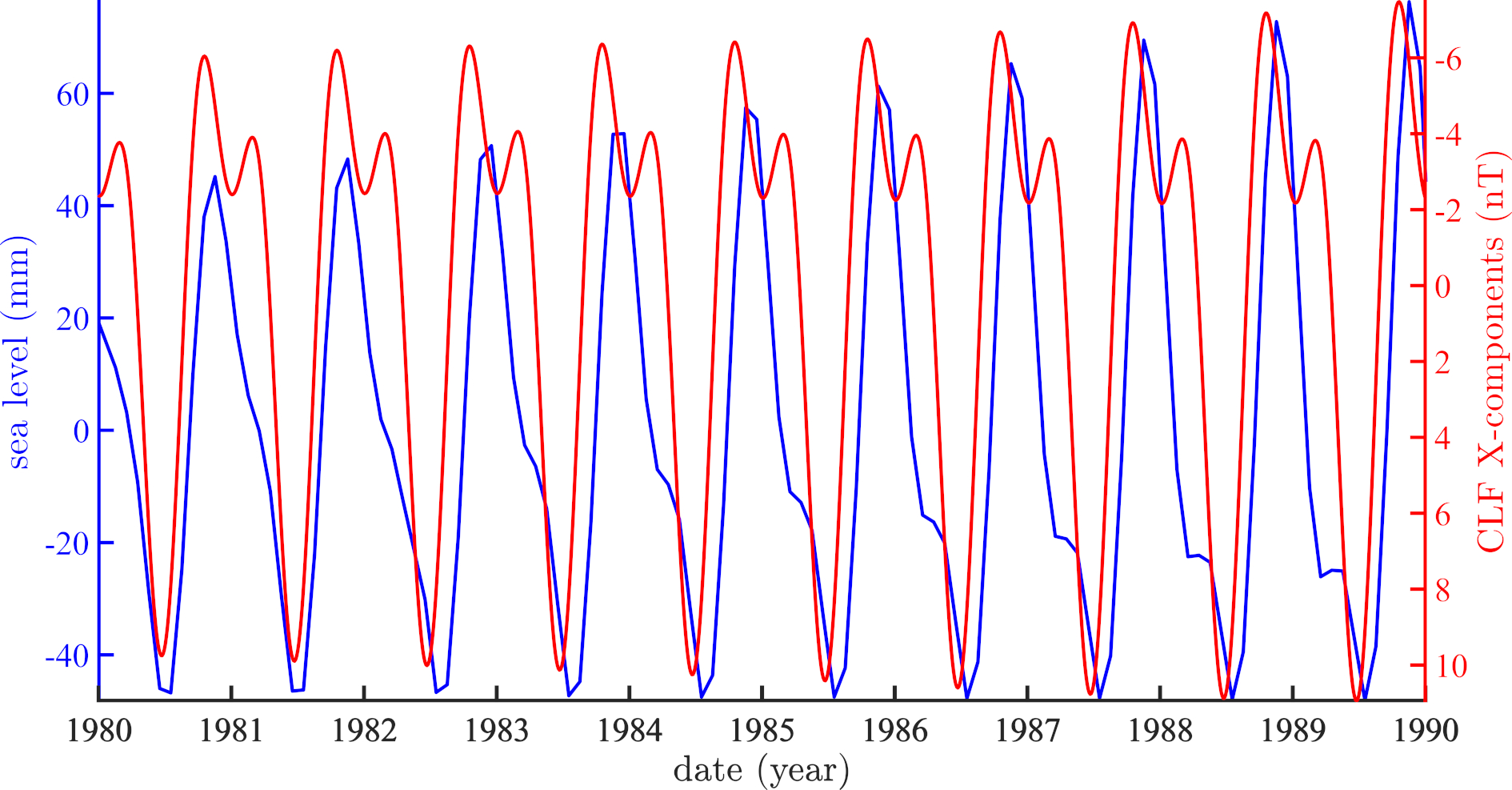} 	
         \caption{Enlargement and superposition of the annual and semi-annual components of the X magnetic component at CLF (red) and sea-level at the Brest tide gauge (blue).}
         \label{fig:05b}
     \end{subfigure}
     \caption{Comparison on the top, between seasonal components of lod (gray curve) and X component of the magnetic field recorded in \textbf{CLF} (red curve); on the bottom same comparison with seasonal component extracted from the Brest tide gauge (blue curve).}
     \label{fig:05}
\end{figure}

	Given their phase and amplitude modulation, the superimposition of the semi-annual and annual components generates fringe patterns (Figure \ref{fig:04a}, \ref{fig:04b}, \ref{fig:04c}). These are better visualized in the enlarged Figure \ref{fig:05a}, \ref{fig:05b} showing the 1980-1990 decade. In Figure \ref{fig:05a} the two components generate a characteristic pattern with two humps that correlate between magnetic field and length of day, with constant phases. In Figure \ref{fig:05b} one of the two humps for sea-level is subdued and looks more like a shifting step. The double hump pattern has already been recognized by \cite{Jault1991}, their Figure 1. Based on Figures \ref{fig:04} and \ref{fig:05}, one can propose that the annual forcing of the double humps is driven by polar motion/rotation. As explained by \Poisson (\cite{Poisson1826}), a moving charged fluid tends to replicate the pattern of the motion; the different processes seem to be in phase and constant.  If these originate from fluid motion tangential to the core, there will be a transfer of moment according to the Liouville-Euler equations. Thus, based on Figures \ref{fig:04} and \ref{fig:05}, one can propose that the annual forcing of the double humps is driven by polar motion/rotation.
 
\begin{figure}[H]
     \centering
     \begin{subfigure}[b]{0.5\textwidth}
         \centering
         \includegraphics[width=1\columnwidth]{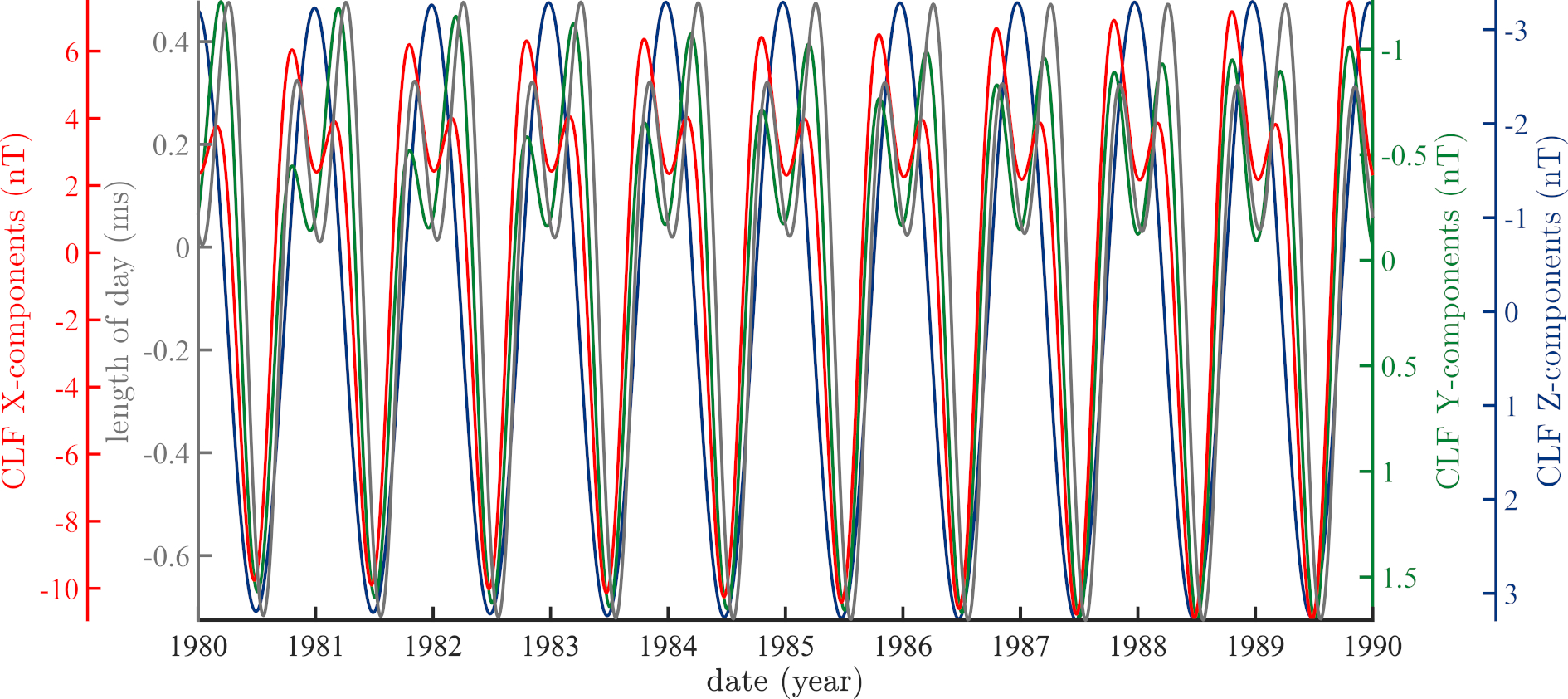}	
         \caption{Comparison of annual plus semi-annual components of all three geomagnetic components X, Y and Z at \textbf{CLF} (X red, Y green, Z blue) with those of \textit{lod} (gray) 1980-1990).}
         \label{fig:06a}
     \end{subfigure}
     \vfill
     \begin{subfigure}[b]{0.5\textwidth}
         \centering
         \includegraphics[width=1\columnwidth]{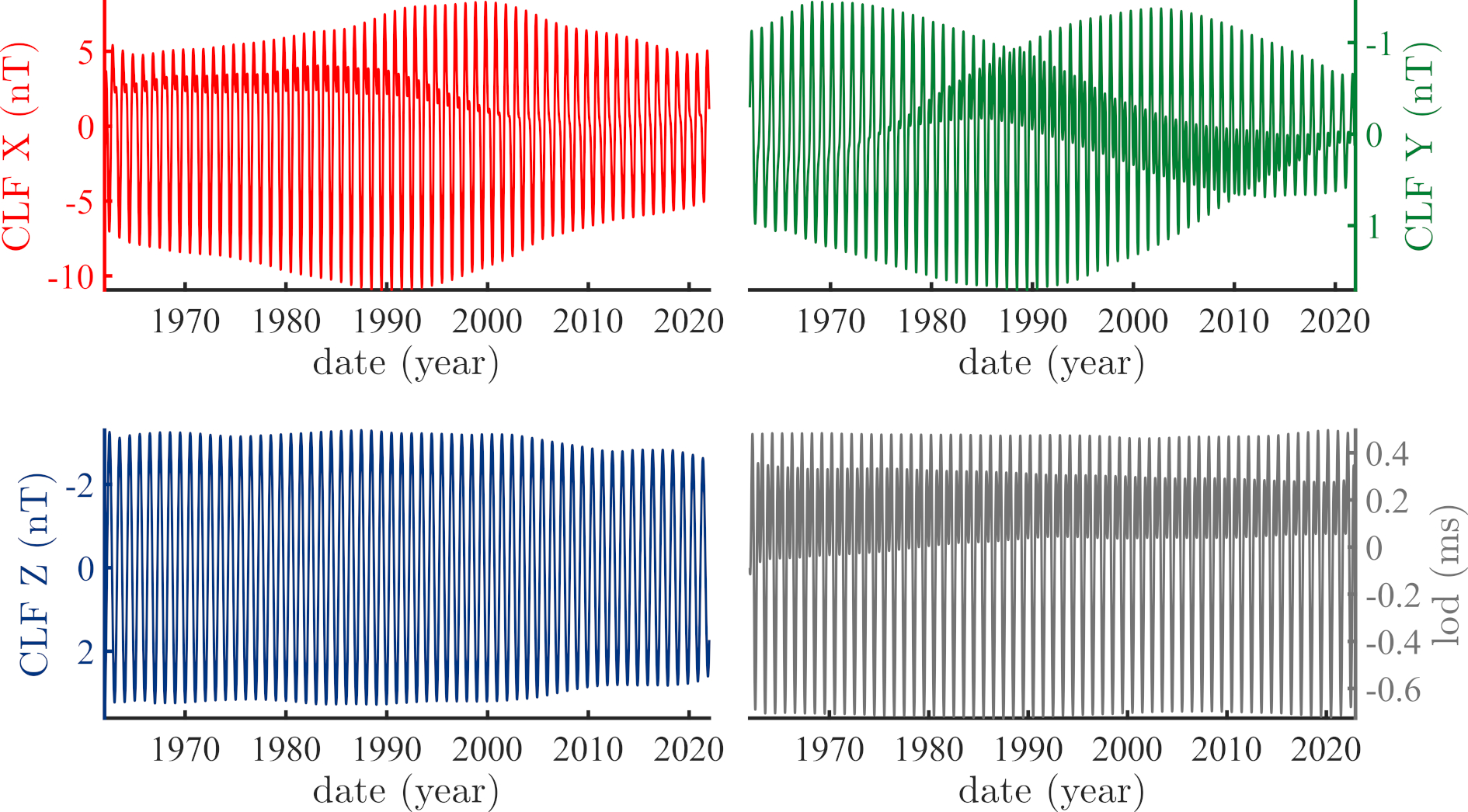} 
         \caption{Annual plus semi-annual components of geomagnetic components X, Y and Z at \textbf{CLF} (X red, Y green, Z blue) and those of \textit{lod} (1962-2022).}
         \label{fig:06b}
     \end{subfigure}
     \caption{Comparison of annual plus semi-annual components of all three geomagnetic components at \textbf{CLF} with those of \textit{lod}.}
     \label{fig:06}
\end{figure}

	We next wish to check whether the same can be said of the other geomagnetic field components at \textbf{CLF}. Figure \ref{fig:06a}, \ref{fig:06b} compares the pattern of \textit{lod} to those of X, Y and Z: Y behaves as X but Z does not have the two humps, only one strong maximum per year (\ie no semi-annual component).

\newpage

\begin{figure}[H]
		\centering{\includegraphics[width=1\columnwidth]{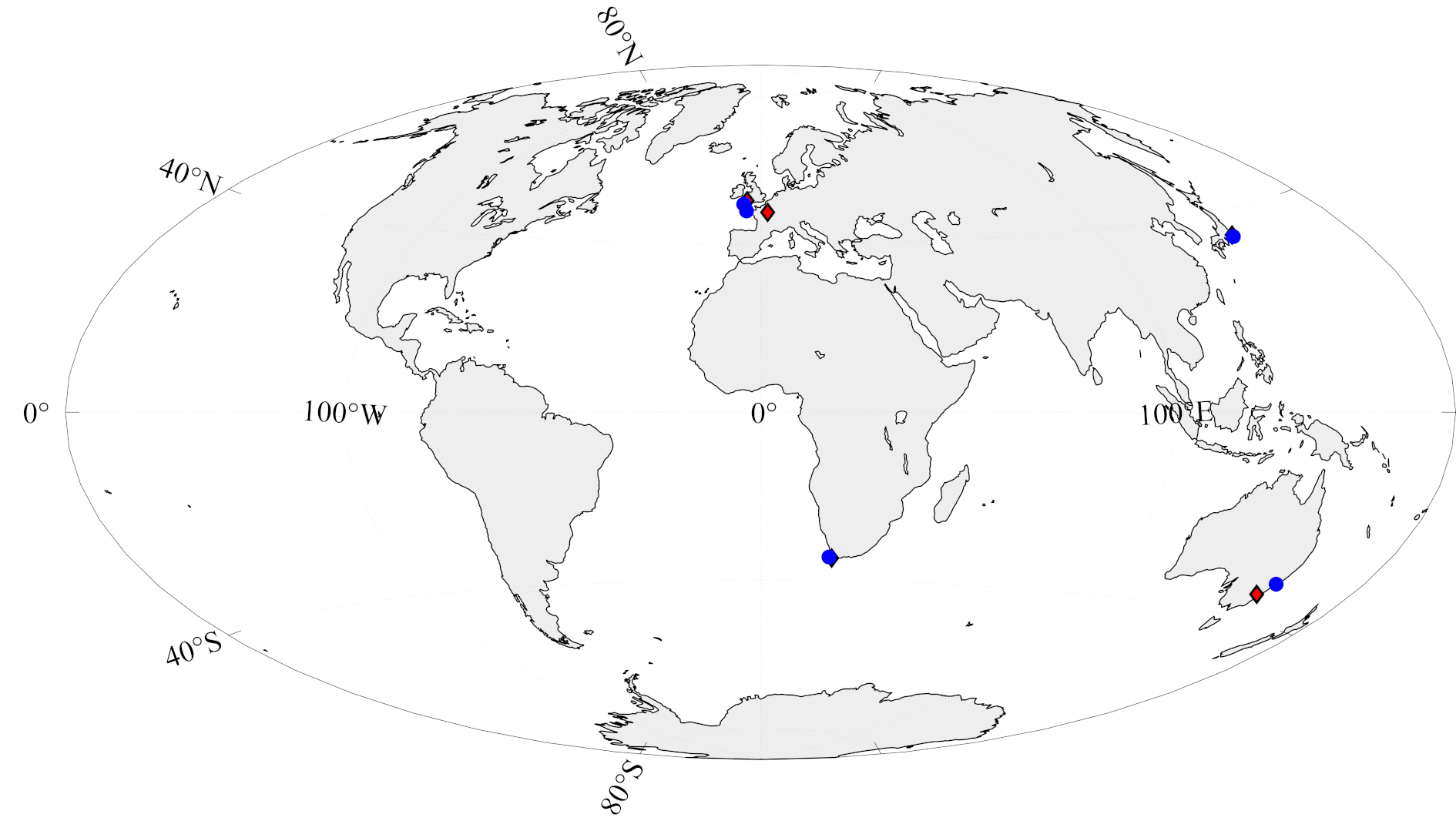}} 	
     	\caption{Associated couples of a magnetic observatory (red diamond) and a tide gauge (blue circles). See Table \ref{tab:01}.}
		\label{fig:07} 	
\end{figure}

	We have attempted to check whether the observations made with the couple "magnetic observatory-tide gauge" at \textbf{CLF} and nearby Brest could be extended to other couples. Unfortunately there are not many such couples, particularly since all data sets should be of sufficient length and quality. We have found five couples listed in Table \ref{tab:01} below and shown on a world map (Figure \ref{fig:07}).	
		  
\begin{table}[H]
\centering
\begin{tabular}{ p{5cm}|p{3cm} }
 \multicolumn{2}{c}{} \\
 \hline
  \hline
	Magnetic observatory & Tide gauge\\
 \hline
 \ & \ \\
Chambon-La-Forêt (\textbf{CLF}, 2.26\degree E, 48.02\degree N)  & Brest (4.49\degree W, 48.38\degree N)\\
Hartland (\textbf{HAD}, 4.48\degree  W, 51\degree N) 			& Newlyn (5.54\degree W, 50.10\degree N)\\
Canberra (\textbf{CNB}, 149.36\degree E, 35.32\degree S) 		& Newcaslte V (151.78\degree E, 32.92\degree S)\\
Hermanus (\textbf{HER}, 19.23\degree W, 34.43\degree S) 		& Simons Bay (18.44\degree E, 34.18\degree S)\\
Kanozan (\textbf{KNZ}, 139.95\degree E, 35.25\degree N) 		&  Mera (139.82°E, 34.91\degree N)\\
\end{tabular}   
    \caption{List of couple "magnetic observatory-tide gauge"}
    \label{tab:01}
\end{table}  
	
\begin{figure}[H]
     \centering
     \begin{subfigure}[b]{0.5\textwidth}
         \centering
         \includegraphics[width=1\columnwidth]{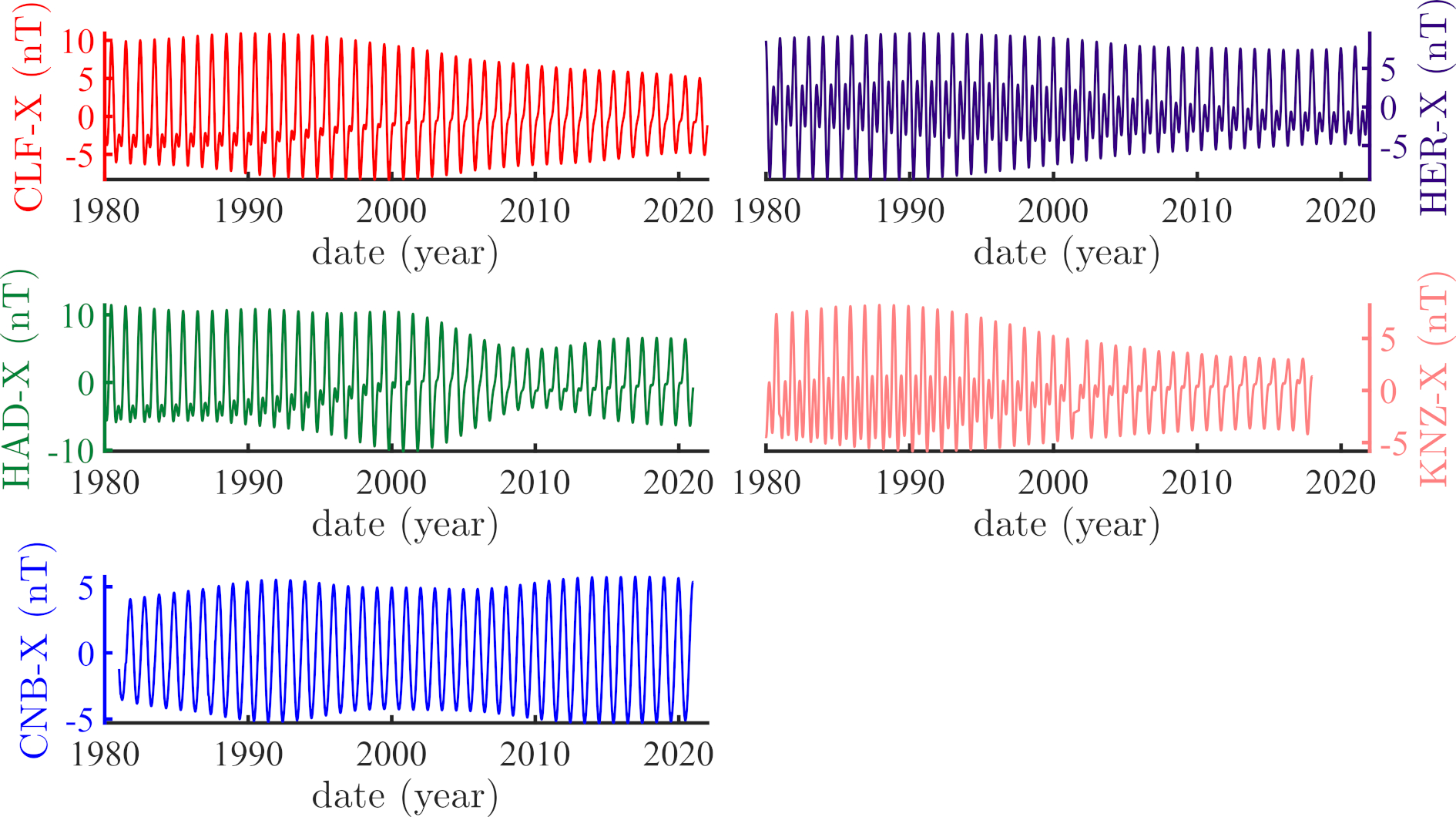}	
         \caption{Forced quasi-cycles associated with the X magnetic component at (from left to right and top to bottom) Chambon-La-Forêt (\textbf{CLF}), Hermanus (\textbf{HER}), Hartland (\textbf{HAD}), Kanozan (\textbf{KNZ}) and Canberra (\textbf{CNB}). See Table \ref{tab:01} and Figure \ref{fig:07}.}
         \label{fig:08a}
     \end{subfigure}
     \vfill
     \begin{subfigure}[b]{0.5\textwidth}
         \centering
         \includegraphics[width=1\columnwidth]{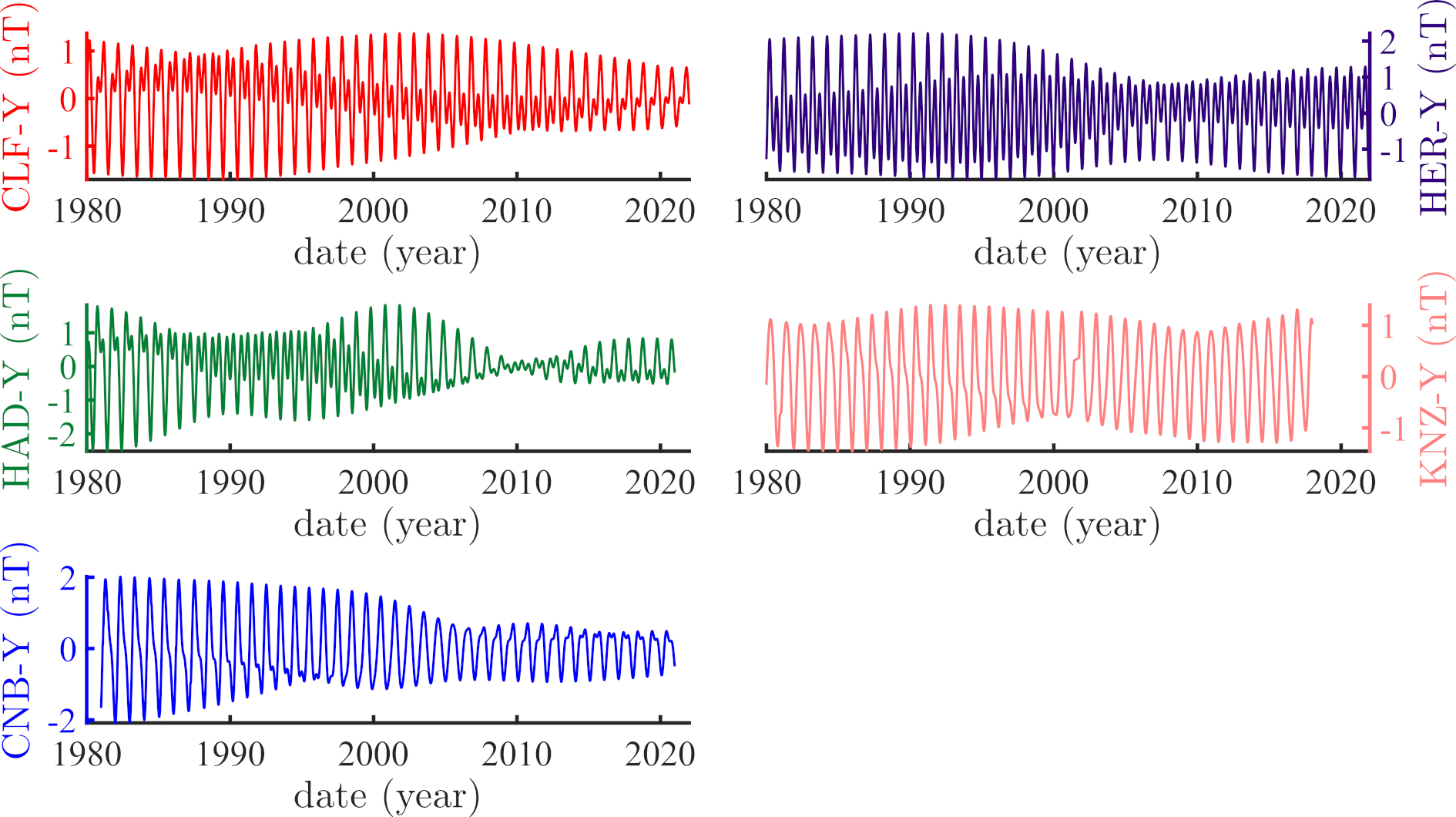} 	
         \caption{Forced quasi-cycles associated with the Y magnetic component at (from left to right and top to bottom) Chambon-La-Forêt (\textbf{CLF}), Hermanus (\textbf{HER}), Hartland (\textbf{HAD}), Kanozan (\textbf{KNZ}) and Canberra (\textbf{CNB}). See Table \ref{tab:01} and Figure \ref{fig:07}.}
         \label{fig:08b}
     \end{subfigure}
       \vfill
     \begin{subfigure}[b]{0.5\textwidth}
         \centering
         \includegraphics[width=1\columnwidth]{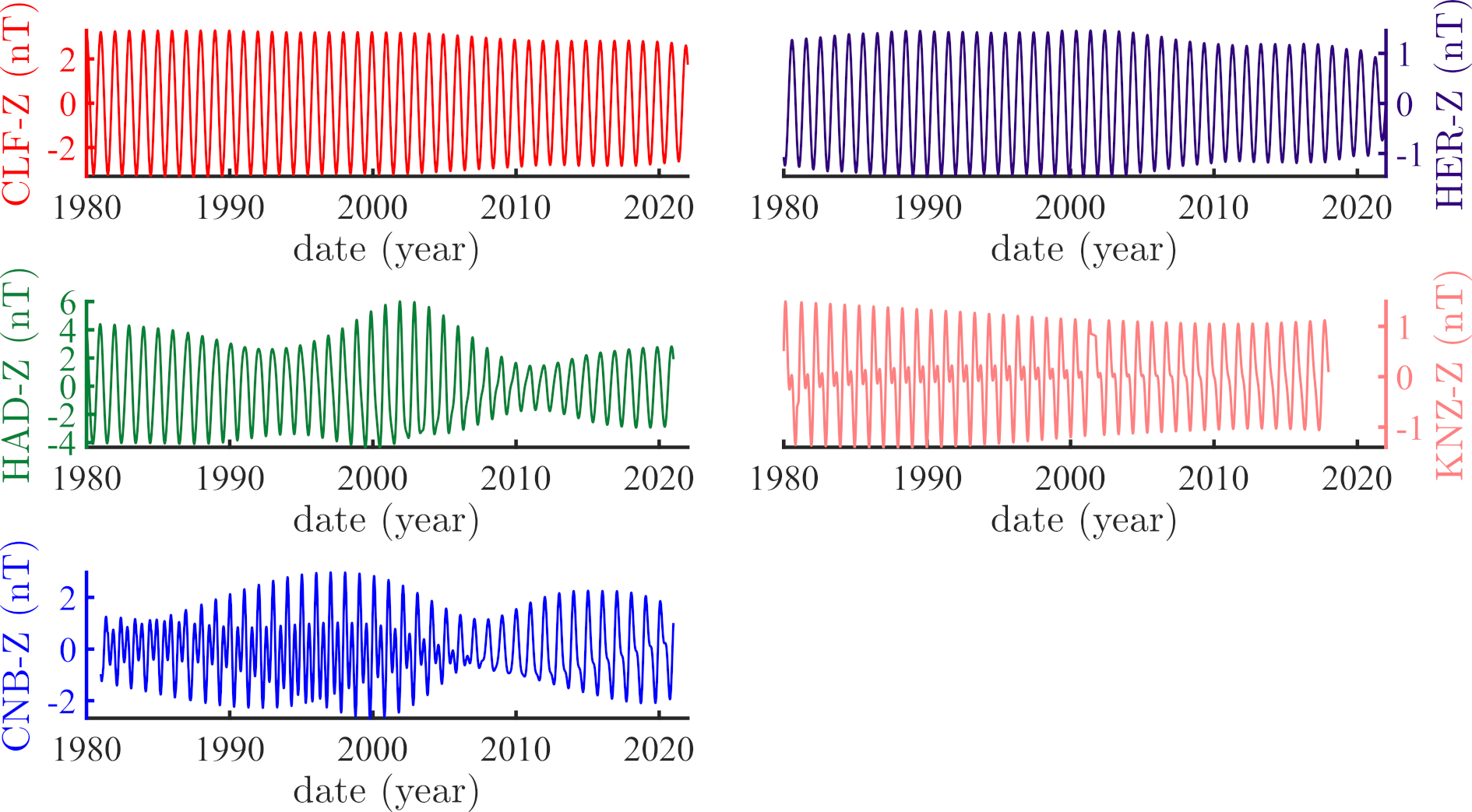} 	
         \caption{Forced quasi-cycles associated with the Z magnetic component at (from left to right and top to bottom) Chambon-La-Forêt (\textbf{CLF}), Hermanus (\textbf{HER}), Hartland (\textbf{HAD}), Kanozan (\textbf{KNZ}) and Canberra (\textbf{CNB}). See Table \ref{tab:01} and Figure \ref{fig:07}.}
         \label{fig:08c}
     \end{subfigure}   
     \caption{Time evolution of annual plus semi-annual components extracted from the 5 observatories listed in Table (\ref{tab:01}).}
     \label{fig:08}
\end{figure}

Figures \ref{fig:08a} to \ref{fig:08c} show the combined annual and semi-annual \textbf{SSA} components of X, Y and Z at the five magnetic observatories. These figures illustrate the different modulation ("wave") patterns associated with annual and semi-annual forcings. There is an ongoing debate as to their origins (\eg \cite{Stewart1882,Stewart1886,Russell1973,Lockwood2020,Lockwood2021}).
\newpage

	Figures \ref{fig:09a} to \ref{fig:09e} allow one to compare the magnetic oscillations for X, Y and Z with the corresponding sea level oscillation, one for each observatory couple. Magnetic components X extracted from the Brest-\textbf{CLF} (Figure \ref{fig:09a}), Simons Bay-\textbf{HER} (Figure \ref{fig:09b}) and Newlyn-\textbf{HAD} (Figure \ref{fig:09c}) couples are in phase opposition with sea-level; the two other field components Y and Z are in phase with sea-level (with a small phase drift over the 40 years of the record). These three couples happen to be located on the same magnetic meridian. The same holds for the Newcastlte-\textbf{CNB} couple (Figure \ref{fig:09e}), with a slightly larger phase drift. 
	
\begin{figure}[H]
     \centering
     \begin{subfigure}[b]{0.5\textwidth}
         \centering
         \includegraphics[width=1\columnwidth]{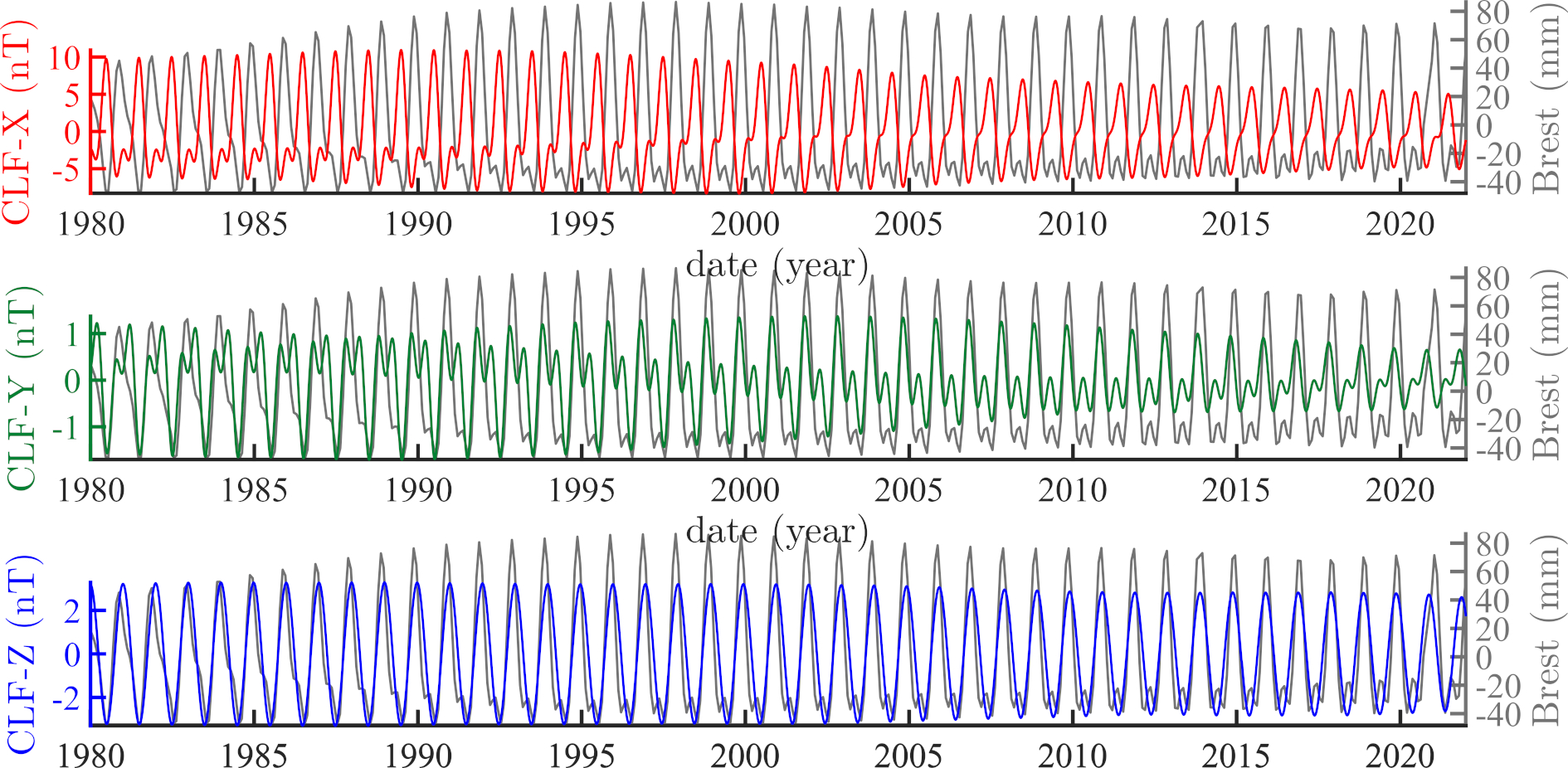}	
         \caption{Comparison between forced components of sea-level in Brest (grey curve) and respective oscillations of magnetic components X (red curve), Y (green curves) and Z (blue curve) in Chambon-La-Forêt.}
         \label{fig:09a}
     \end{subfigure}
     \vfill
     \begin{subfigure}[b]{0.5\textwidth}
         \centering
         \includegraphics[width=1\columnwidth]{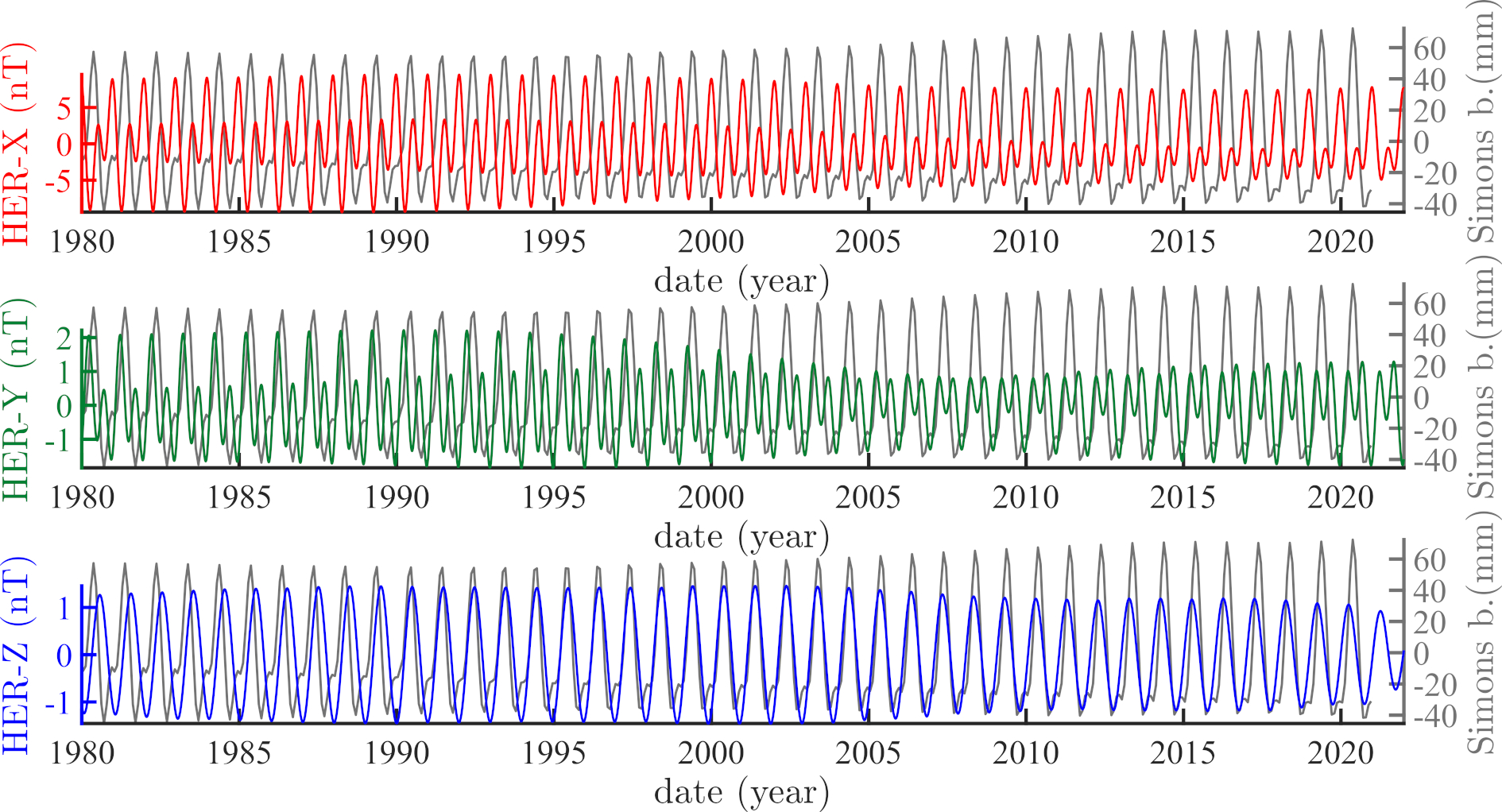} 	
         \caption{Same as Figure \ref{fig:09a} for the tide gauge/magnetic observatory couple Simons Bay/Hermanus.}
         \label{fig:09b}
     \end{subfigure}
       \vfill
     \begin{subfigure}[b]{0.5\textwidth}
         \centering
         \includegraphics[width=1\columnwidth]{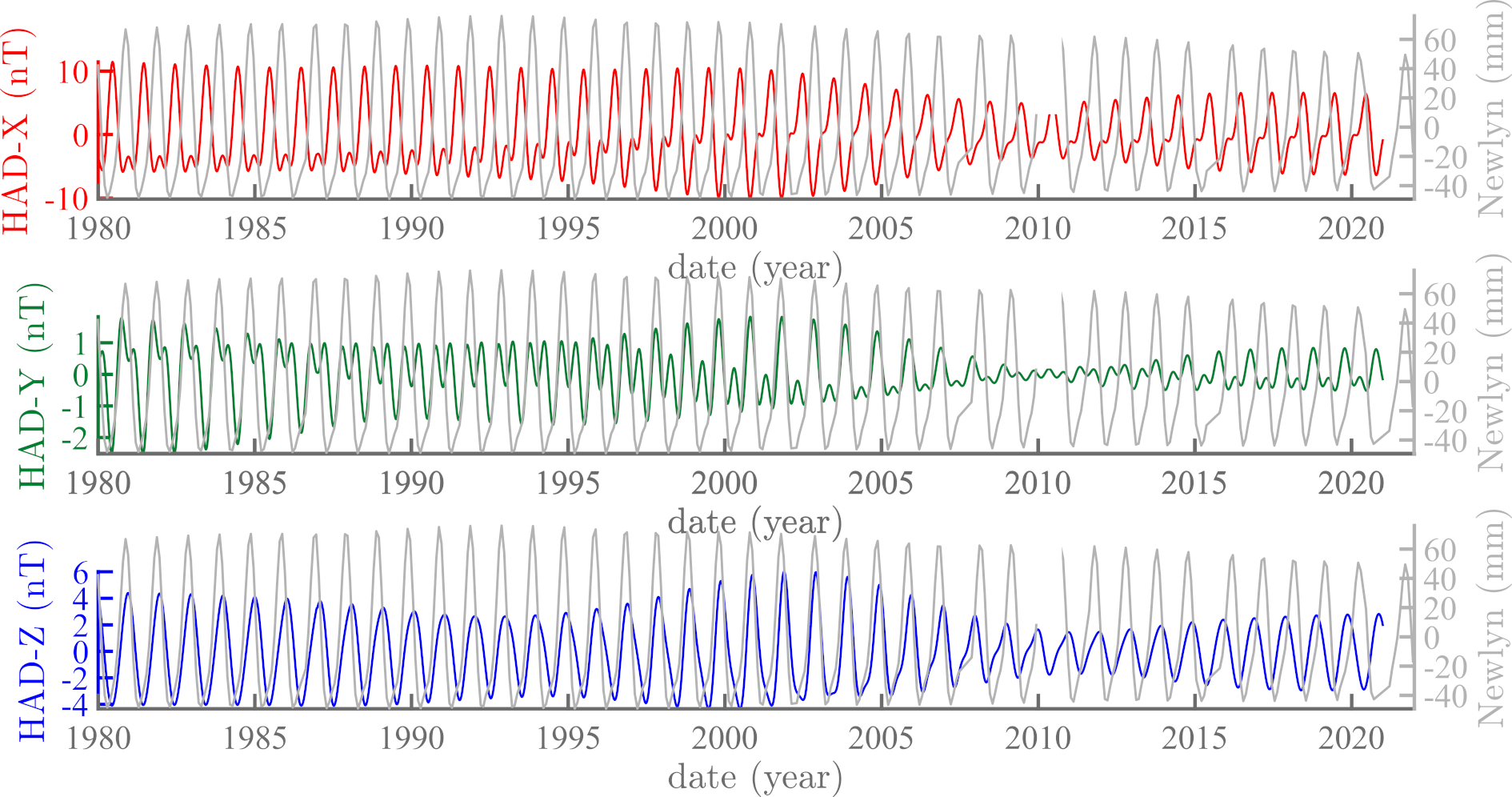} 	
         \caption{Same as Figure \ref{fig:09a} for the tide gauge/magnetic observatory couple Newlyn/Hartland. }
         \label{fig:09c}
     \end{subfigure}   
\end{figure}	
\newpage
\begin{figure}[H]\ContinuedFloat
    \centering
     \begin{subfigure}[b]{0.5\textwidth}
         \centering
         \includegraphics[width=1\columnwidth]{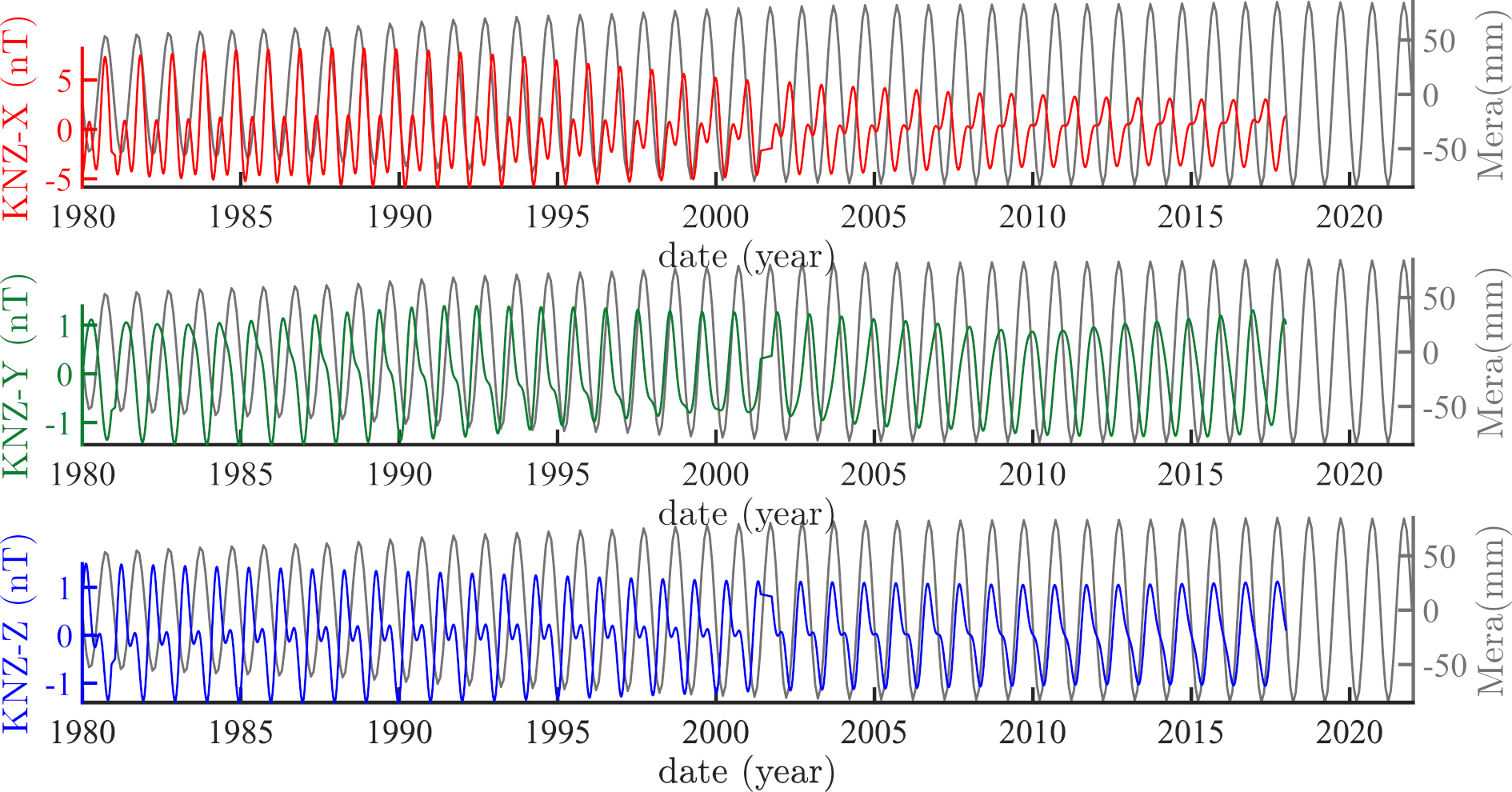} 	
         \caption{Same as Figure \ref{fig:09a} for the tide gauge/magnetic observatory couple Mera/Kanozan.}
         \label{fig:09d}
     \end{subfigure}  
            \vfill
     \begin{subfigure}[b]{0.5\textwidth}
         \centering
         \includegraphics[width=1\columnwidth]{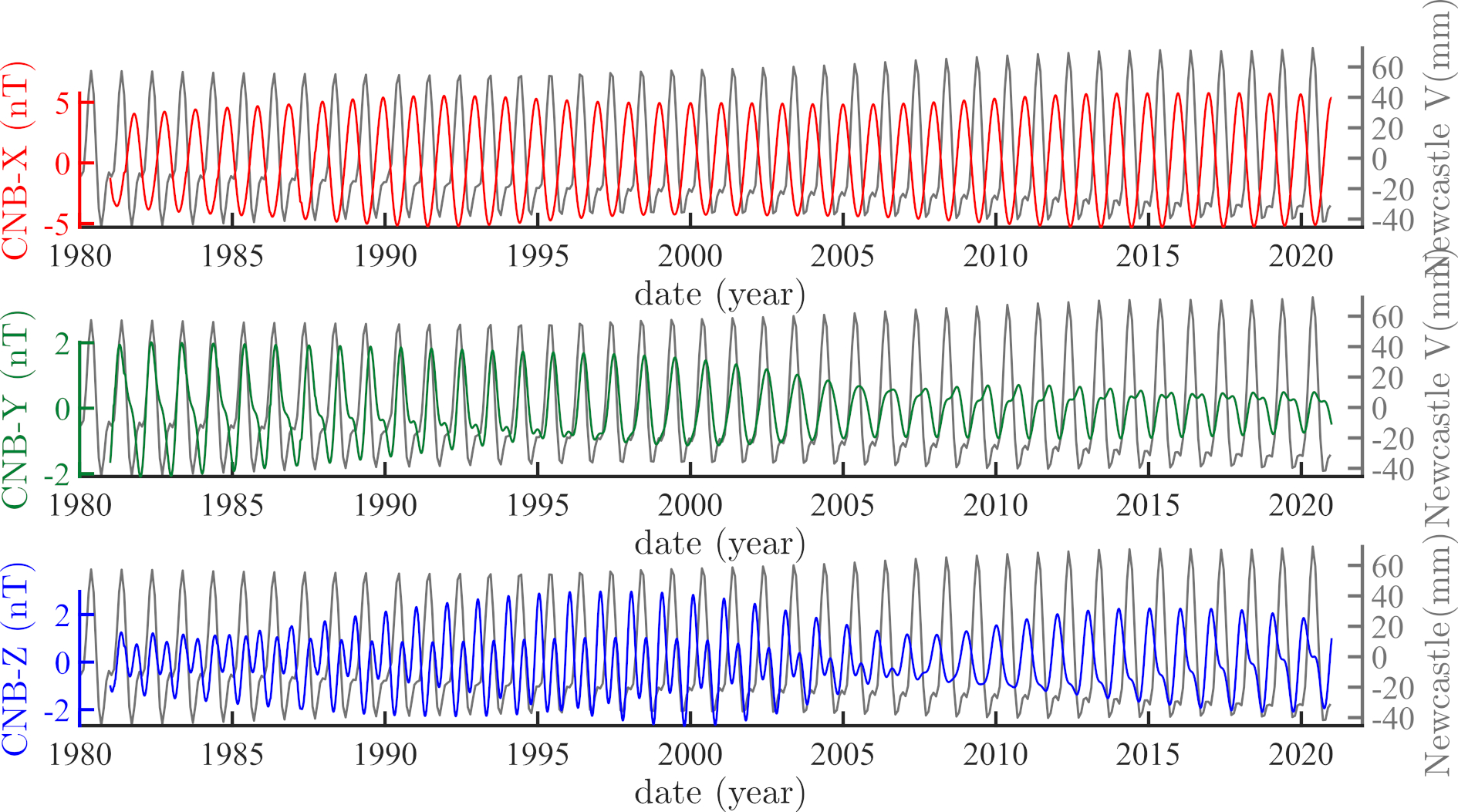} 	
         \caption{Same as Figure \ref{fig:09a} for the tide gauge/magnetic observatory couple Newcastle/Canberra. }
         \label{fig:09e}
     \end{subfigure}  
     \caption{Comparison between forced components extracted from each couples observatory/tide gauge according listed on the Table (\ref{tab:01})}
     \label{fig:09}
\end{figure}		
	
	For the Japanese couple Mera/\textbf{KNZ} (Figure \ref{fig:09d}), X is in phase with sea-level in 1980, when Y and Z are in quadrature. After 40 years of slow drift, Z is in phase, X and Y in quadrature. Finally, for the Australian couple Newcastle/\textbf{CNB} (Figure \ref{fig:09e}), X is in phase opposition, Y in phase and Z in quadrature in 1980 and the three drift respectively to opposition, quadrature and opposition in 2020. We note that in Hartland Z does not have a semi-annual component (Figure \ref{fig:09c}), and it is quite small in Hermanus (Figure \ref{fig:09b}). We have tested tens of potential tide gauge/magnetic observatory couples whose results are not good enough to be reported; we just note that some 80\% of them have no semi-annual Z. 
	
	The comparisons made in Figures \ref{fig:09} suggest strongly to us (though, granted, they do not prove) a link between annual variations of sea-level and the magnetic field. The correlations are of better quality than we might have expected, despite differences in topography and geography in the vicinity of the gauges. The same can be said of the (\textbf{SSA} determined) trends. We are in a position to strengthen our physical understanding of this link.
	
	In section \ref{sec2} we have seen that if a measured magnetic field has its origin in the uniform motion of charges in rotation about our planet’s symmetry axis, the quantity and quality of charges cannot vary with time (or the dipole would vanish). Then there is a link between the magnetic and mechanical moments. It is expressed by equation (\ref{eq:08a}):
\begin{equation*}	
			\mathfrak{m} = \dfrac{e}{2mc} \sum \textbf{r} \times \textbf{p} = \dfrac{e}{2mc}\mathcal{M}
\end{equation*}
	
	Sea-level and magnetic variations are linked through polar motion (\eg \cite{LeMouel1984}, and section \ref{sec2}), here the length of day. Polar motion is forced by the Earth's revolution about the Sun. If the field is constant and the link exists, the ratio	$\dfrac{e}{2mc}$ must be constant. The (geo-)physical consequences generated by these moments should be to first order the same for all torques around the globe, and the variations in amplitude of the geophysical phenomena involved should be proportional. 
	
	In Table \ref{tab:02} below, we evaluate the amplitudes of the \textbf{SSA} annual components of sea level and magnetic components and their ratios for the five couples of stations of Figure \ref{fig:07}. In the 1980-2022 period, these ratios are the same at 7mm/nT for \textbf{CLF}, \textbf{HAD} and \textbf{HER}, almost the same at 8 mm/nT for \textbf{CNB} and not so different at 10 mm/nT for \textbf{KNZ}. Given the complexity of sea-level physics and geomagnetism, there was no a priori reason why the ratios should be constant, unless equation (08) holds, which seems to be the case. This result vindicates \Poisson 's approach (\cite{Poisson1826}): fluid motions in the core are similar to those at the surface; because they are charged, the motifs of variations in the Earth’s magnetic field are those of sea-surface, and atmosphere. 
\begin{table}[H]
\centering
\begin{tabular}{ p{3cm}|p{3cm}|p{2cm} }
 \multicolumn{3}{c}{} \\
 \hline
  \hline
	couple observatory-tide gauge & ratio sea level / magnetic component & order of magnitude\\
 \hline
 \ & \ & \ \\
 \textbf{CLF}/Brest, between 1980 and 2000        & $\sim$100 mm / 15 nT & $\sim$7 mm/nT \\
 \textbf{HAD}/Newlyn, between 1980 and 2005       & $\sim$100 mm / 15 nT & $\sim$7 mm/nT \\
 \textbf{CNB}/Newcaste V, between 1980 and 2022   & $\sim$ 80 mm / 10 nT & $\sim$8 mm/nT \\
 \textbf{HER}/Simons bay, between 1980 and 2000   & $\sim$100 mm / 14 nT & $\sim$7 mm/nT \\
 \textbf{KNZ}/Mera, between 1980 and 1995         & $\sim$140 mm / 14 nT & $\sim$10 mm/nT \\
\end{tabular}   
    \caption{List of couple "magnetic observatory-tide gauge"}
    \label{tab:02}
\end{table}  	
		
	\subsection{On the 11-yr cycle and the magnetic field}\label{4-3}
	The 11-yr cycle is one of the better known variations of the Earth’s magnetic field. It is considered as the same \blue{Schwabe} (\cite{Schwabe1844}) cycle that is found in sunspot numbers. In the minds and logic of \Laplace (\cite{Laplace1799}), \Lagrange (\cite{Lagrange1788}) and \Poisson (\cite{Poisson1826}), planets are responsible, through exchanges in moment, for a number of astrophysical and geophysical phenomena. The origin of this 11-yr cycle has likely its source in the revolution and moment of Jupiter. This moment is directly connected to variations of distances to Jupiter in the solar system. Jupiter exerts the largest torque of the 8 planets (and also larger than the Sun’s, that is motionless at the time scales we are interested in \blue{Lopes et \textit{al.}} \cite{Lopes2022c}). This torque acts on (or modulates, or forces) sunspots (\eg \cite{Courtillot2021}) as well as on polar rotation (\eg \cite{Lopes2017,Lopes2021}).
	
		In what we will call the \Poisson - \LeMouel 's paradigm, the fluid’s rotation generates the field; this same rotation must then force solar activity in the form of sunspots. In the case of fluid mechanics, and as shown in \blue{Courtillot and Le Mouël} (\cite{Courtillot1988}) figure 45 and in \blue{Le Mouël et \textit{al.}} (\cite{LeMouel2019a,LeMouel2020}), a law of natural turbulence appears, a \cite{Kolmogorov1941} power law with exponent -5/3. This law is found in sunspots as well as in variations of geomagnetic intensity (the latter since as far back as at least 1 Myr, see \cite{Courtillot1988}). The angular coordinate $\theta$ and $\dfrac{d\psi}{dt}$,  that define the location of the pole of rotation on the sphere are the solutions of a system of differential equations, one for $\theta$, that describes pole motion, the other for $\dfrac{d\psi}{dt}$ that describes length of day. One is the derivative of the other (see \cite{Lopes2022b}). As is the case for all derivative operators, it amplifies high frequency components. \blue{Le Mouël et \textit{al.}} (\cite{LeMouel2019b}) have shown that the 11-yr component is a major component of length of day, whereas \blue{Lopes et \textit{al.}} (\cite{Lopes2017}) have shown that it was a minor one of polar motion.
\begin{figure}[H]
		\centering{\includegraphics[width=1\columnwidth]{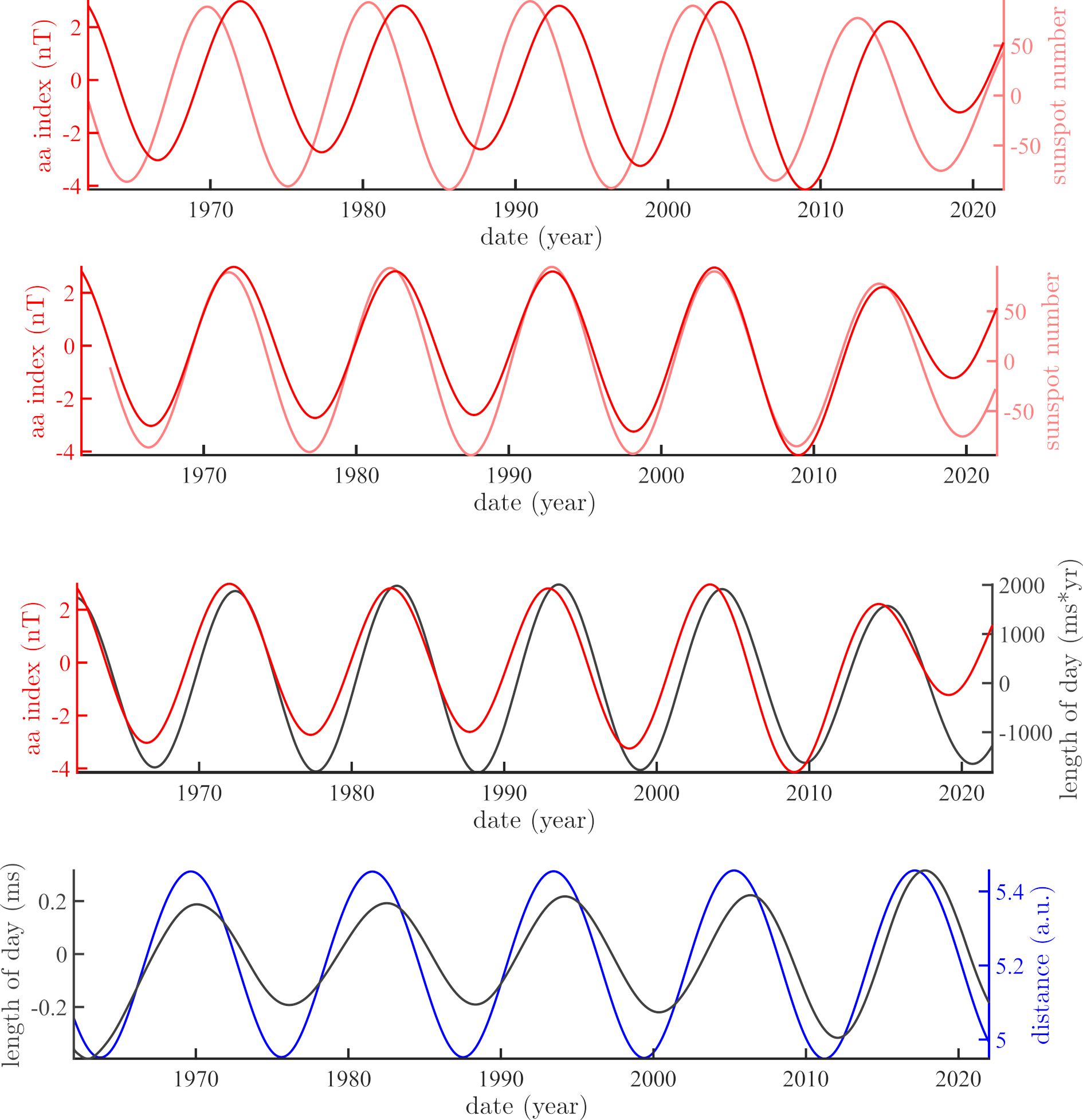}} 	
     	\caption{Eleven year quasi-cycles extracted by SSA from the geomagnetic index aa (red; top 3 rows), the sunspot series (pink; top 2 rows), the length of day (black; bottom 2 rows) and the ephemerids of Jupiter (blue; bottom row marked as distance of Earth from Jupiter).}
		\label{fig:10} 	
\end{figure}
		
	In order to retain the homogeneity of our data sets and their temporal resolution, we have integrated the 11-yr quasi-cycle extracted from length of day by \textbf{SSA} (Figure \ref{fig:10}, black curve, bottom two rows). In the top row of  Figure \ref{fig:10}, we have superimposed the 11-yr component from sunspots (pink) with that of \textit{aa}. The two curves appear to be in quadrature: this is checked by offsetting the Schwabe cycle forward by exactly 11/4 yr (second row). The explanation for this observation is the following: the torque exerted by Jupiter acts directly on sunspots, while the aa index is the difference between two antipodal observatories. Thus the \textit{aa} index is a derivative operator. This is likely why the 11-yr cycle is prominent in \textit{aa} but minor in the X, Y and Z components. The same accounts for the phases of aa and lod: we integrate the 11-yr component of lod (black curve in 3rd row of  Figure \ref{fig:10}) and see it is in phase with aa. And finally, according to \cite{Lagrange1788}, Jupiter does act on the Earth’s rotation, as shown by \blue{Lopes et \textit{al.}} (\cite{Lopes2021}) and the Jupiter-Earth distance (blue curve bottom row in Figure \ref{fig:10}). This last result provides a good illustration of equation (\ref{eq:08d}):
\begin{equation*}
		\dfrac{d\mathcal{M}}{dt} = -\Omega \times \mathcal{M}.
\end{equation*} 

Polar motion as well as length of day are linked to $\Omega$ (\eg \blue{Lambeck} \cite{Lambeck2005}, section 03). We have also seen in theory (and checked in the example above) that the variation in the magnetic field is linked to moment $\mathcal{M}$. We could say that magnetic field components (X,Y,Z) are to polar motion ($m_1$,$m_2$) what \textit{aa} indices are to length of day. The link between sections \ref{sec2} and \ref{sec3} above is the length of day.
		
	\subsection{On the international geomagnetic reference field}\label{4-4}
	Since the work of \Gauss (\cite{Gauss1837}), the decomposition of the geomagnetic field into spherical harmonics has become normal practice (“routine”). We recall however as a caveat the fact that an electric field or a gravity field can be decomposed in SH because these fields are constant and their elementary sources all have the same sign. Such is not the case with a magnetic field, unless it is constant also (sections 2 and 3). In principle, the SH decomposition of a magnetic field does not have physical significance. Nevertheless a spherical harmonic decomposition of the International Geomagnetic Reference Field (\textbf{IGRF}) is published every five years (\cite{Alken2021}). Given the fact that there is no magnetic monopole, the first source term (aka "\Gauss coefficient") is the axial dipole $g_{1,0}$, an imaginary source at the center of the Earth. The other terms of the \blue{Fourier} expansion on the base functions $\textrm{cos}$ and $\textrm{sin}$ are written as $g_{l,m}$ and $h_{l,m}$.	
	
\begin{figure}[H]
     \centering
     \begin{subfigure}[b]{0.5\textwidth}
         \centering
         \includegraphics[width=1\columnwidth]{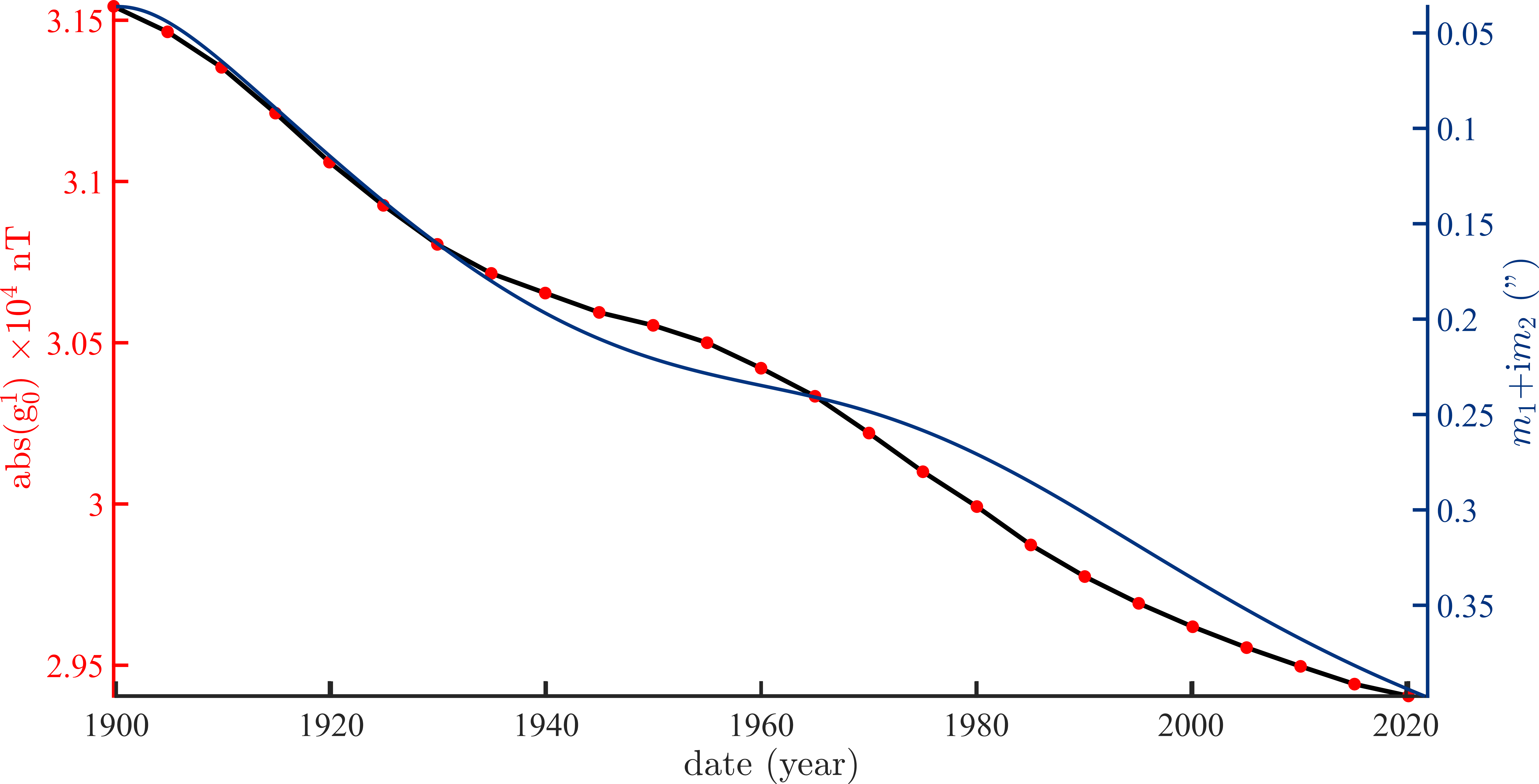}	
         \caption{Red dots = \textbf{IGRF} $g_{1}^{0}$ every 5 years since 1900 \cite{Alken2021} and interpolation as the black curve. Blue curve = \textbf{SSA} first component \ie trend of polar motion, that is the Markowitz-Stoyko drift.}
         \label{fig:11a}
     \end{subfigure}
     \vfill
     \begin{subfigure}[b]{0.5\textwidth}
         \centering
         \includegraphics[width=1\columnwidth]{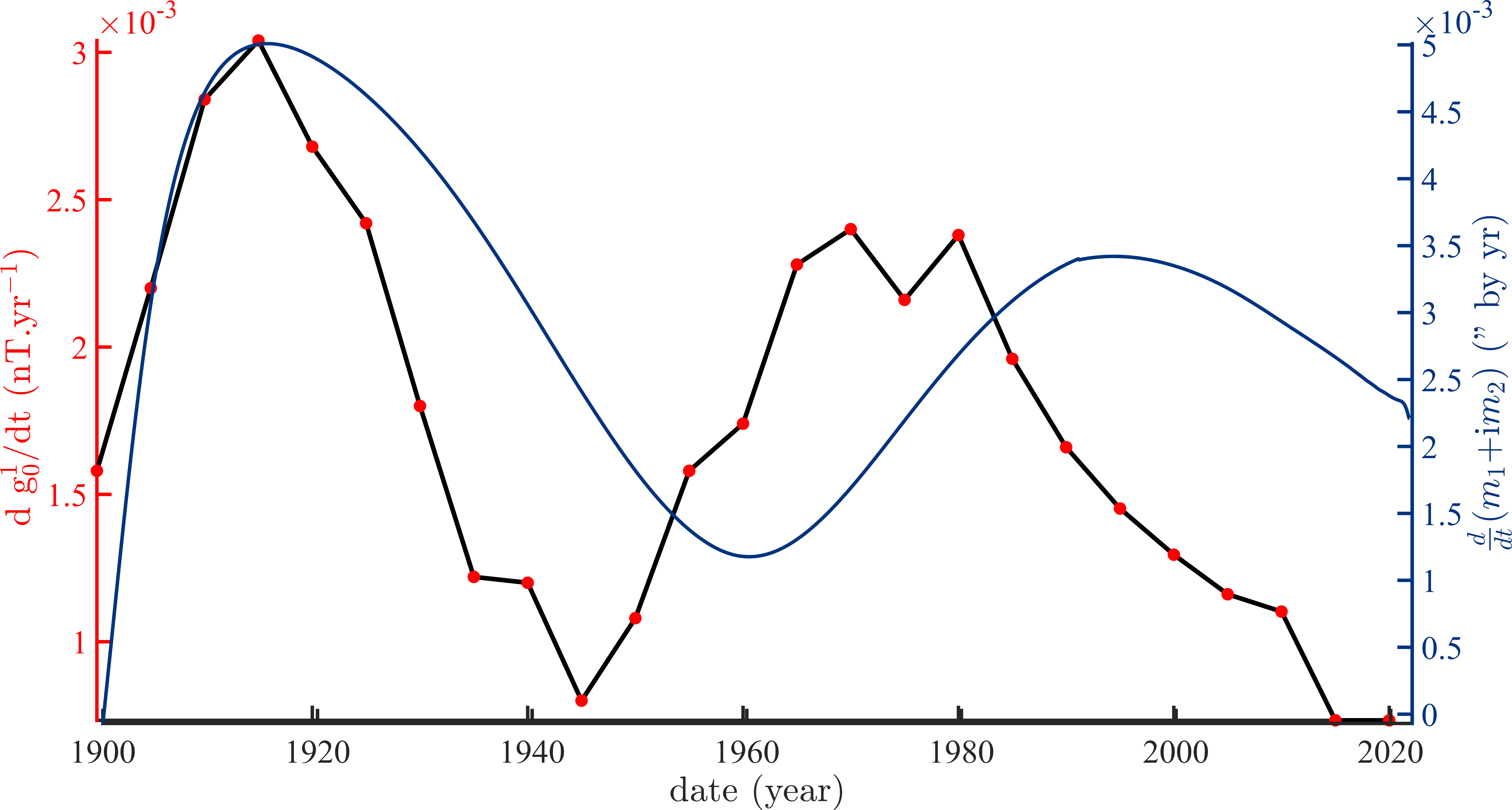} 	
         \caption{First time derivatives of the two curves shown in figure \ref{fig:11a}}
         \label{fig:11b}
     \end{subfigure}
       \vfill
     \begin{subfigure}[b]{0.5\textwidth}
         \centering
         \includegraphics[width=1\columnwidth]{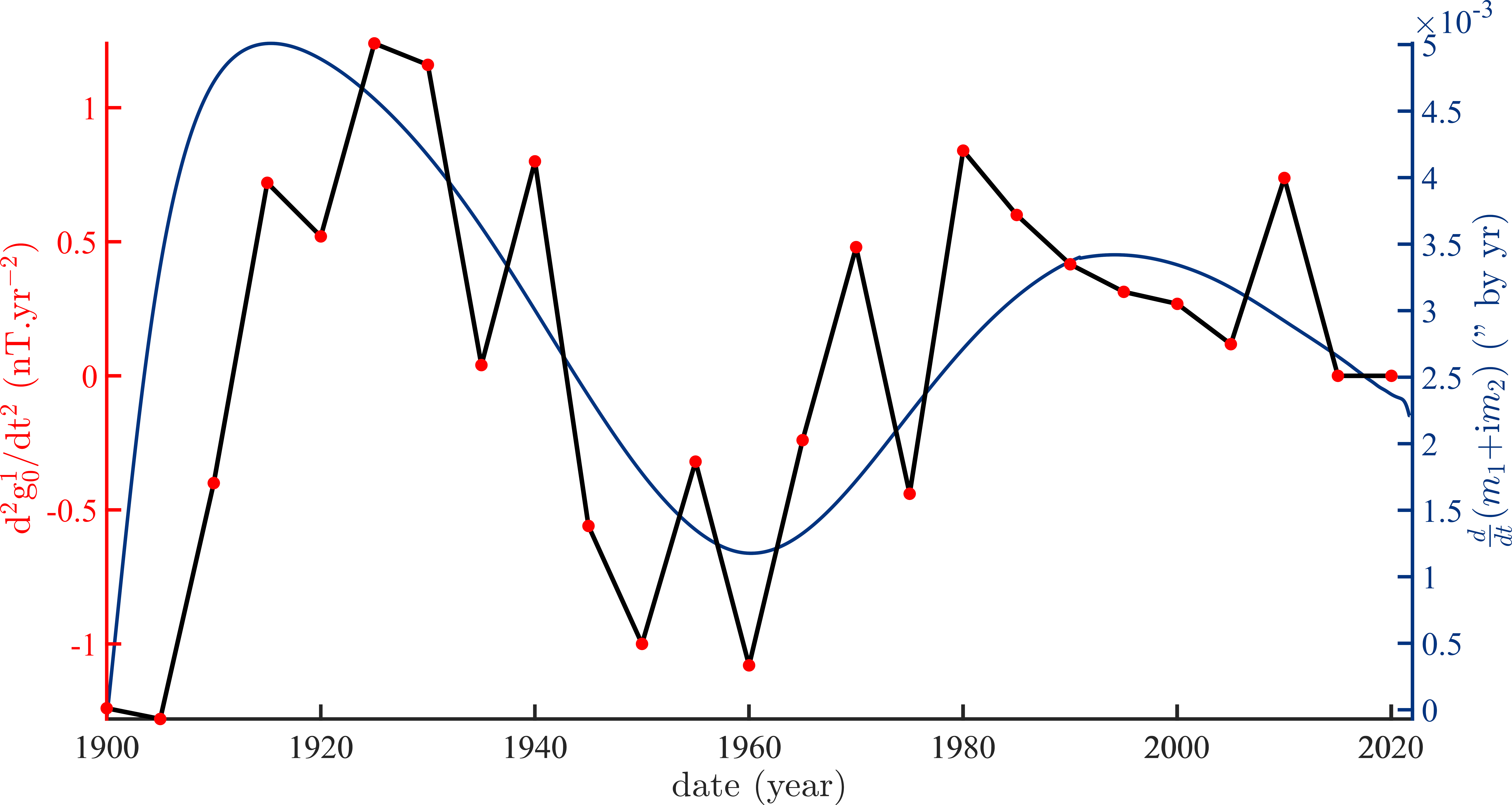} 	
         \caption{ Second derivative in time of the curve corresponding to the coefficients$g_{1}^{0}$ versus the first derivative of the polar motion. The two curves are in phase possibly due to causality.}
         \label{fig:11c}
     \end{subfigure}
     \label{fig:11}
     \caption{Comparison between (a) the evolution of $g_{1,0}$ from (\cite{Alken2021}) and the Markowitz-Stoyko drift and (b-c) their corresponding time derivatives.}   
\end{figure}

	Figure \ref{fig:11a} shows the monotonous decay of the \textbf{IGRF} axial dipole $g_{1,0}$ since 1900. With \Poisson 's (\cite{Poisson1826}) and \LeMouel 's (\cite{LeMouel1984}) hypothesis in mind, and given some of the results in the previous sections of this paper (similar behavior of the annual \textbf{SSA} components of sea level, rotation axis and magnetic field), it is natural to compare the behavior of the intensity of the \textbf{IGRF} dipole with polar motion ($m_1$,$m_2$) or the equivalent parameter $\theta$ of \cite{Laplace1799}. This is done in Figure \ref{fig:11a}.
	
	The time variations of $g_{1}^{0}$ and $\theta$ are indeed very close.  The first derivative of polar motion is in quadrature with the first derivative of $g_{1}^{0}$ (Figure \ref{fig:11b}) and in phase (opposition) with the second derivative of $g_{1}^{0}$ (Figure \ref{fig:11c}).The variation of dipole intensity follows the variation of the angle $\theta$ between the rotation axis of the planet and that of the dipole.  Another example is given by the similarities between magnetic declination in Paris and rotation axis, and between their derivatives (Figures  \ref{fig:03a} and \ref{fig:03b}).
	
	If the field is constant, we have seen that there is a link between the magnetic moment $\mathfrak{m}$ and the kinetic moment $\mathcal{M}$
\begin{equation*}
	\mathfrak{m} = \dfrac{e}{2mc} \sum \textbf{r} \times \textbf{p} = \dfrac{e}{2mc}\mathcal{M}		
\end{equation*}
\begin{equation*}
	\dfrac{d\mathcal{M}}{dt} = -\Omega \times \mathcal{M}
\end{equation*}
		
	We have checked this relation, using observations (Table \ref{tab:02})  We have also shown that, in the case of a rotating system of charges about the axis $Z$ (with velocity $\Omega$), this relation becomes:
\begin{equation*}
	\boldsymbol\Omega = \dfrac{e}{2mc}\textbf{H} 	
\end{equation*}	

	This does express the link between the variations of the Earth’s rotation and the magnetic field \textbf{H}. Since $\Omega$ is connected to $m_{1}$ and $m_{2}$ through the Liouville-Euler equations, then \textbf{H} is also connected to $m_{1}$ and $m_{2}$.  But the key coordinate is $\dot{m}_{3}$ that is the length of day. This is the reason why the second derivation of to $g_{1,0}$ agrees better with the \blue{Markowitz-Stoyko} drift, which is as we have seen the second derivative of \textit{lod}.

\section{Conclusion}\label{sec5}	
	Our aim in this paper was to assess \Poisson 's pioneering work (\cite{Poisson1826}) on the nature and origin of the Earth’s magnetic field. These works started with the scientist’s recognition that the field had to be constant, which may seem awkward to the modern physicist. \Gauss (\cite{Gauss1837}) also considered that the magnetic field was constant.
	
	\LeMouel \cite{LeMouel1984} observed that there was a strong morphological link between the secular variation of the field ($\dfrac{d \textbf{H}}{dt}$) and the drift of the rotation pole (polar motion $m_{1}$ and $m_{2}$), and another link between the forced oscillations of the magnetic field \textbf{H} and the length of day $\dot{m}_{3}$ (\ie rotation velocity). Based on these observational facts, \cite{LeMouel1984} proposed a model, a rotating cylinder, that was in many ways identical to \Poisson’s sphere, though he did not state whether the field was constant or not. Poisson’s proof that indeed the magnetic field had to be constant involved theoretical physics and mathematics but only a limited set of observations. Based on a large number of observations, \LeMouel (\cite{LeMouel1984}) and collaborators built quasi “experimentally” the same theory as \Poisson in a suite of papers. These papers are used significantly in the literature; it is not clear whether they are compatible with the complex dynamo models that are often preferred today.
	
	We have attempted in this paper to return to the original sources and to reconstruct the development of geomagnetism, using “modern” language. Starting from Maxwell’s equations, we have derived the equations for the electrostatic and magnetostatic fields (section \ref{sec2}). We come to the same equations as Poisson with an important difference: Poisson chose the Lagrangian approach to gravity (\cite{Legendre1785,Lagrange1788,Laplace1799}), to formally derive the equations for the magnetic field (\cf \cite{Lopes2023}). To scientists of this epoch, there was no difficulty in reasoning in magnetism in the same “classical” way one reasoned in gravity (as a side note, this led to a beautiful piece by \blue{Heaviside} \cite{Heaviside1893}, on the deep analogy between magnetism and gravity).
	
	We have pursued a more classical approach, and in section \ref{sec2} we propose a synthesis with comments of the equations of electromagnetism that can be found in most physics graduate textbooks. One of the points we wish to emphasize is that the magnetic field need not derive from a scalar potential to be developed into spherical harmonics. However, most geomagnetists do make this hypothesis, invoking Stokes’s theorem: since magnetic measurements are made at the surface where almost no charge circulates, one can assume that the field derives from a scalar, not a vector potential. It is more physical, hence logical to obtain the spherical harmonics from the electrostatic field, then using the vector potential (equation \ref{eq:06c}) to return to their expression for the magnetic field. 
	
	We have not implemented this last step because one can perform a decomposition in spherical harmonics and write about multi-poles if and only if the motions of charges are finite and uniform, two conditions that are not met in a dynamic field.  From this, we have drawn a number of consequences for the magnetic field, the main one being that the magnetic moment of the charges that generate the field and their mechanical moment (thus the motion of the rotation pole) are linked by Larmor’s relation. This is in agreement with the theoretical works of \Laplace , \Poisson (\cite{Laplace1799,Poisson1826}) and \cite{LeMouel1984}. A magnetic field can be written on a basis of spherical harmonics only if this field is constant. Section \ref{sec3} has dealt with the intrinsic theoretical consequences of a constant dipolar field. 
	
	From the ideas presented in sections \ref{sec2} and \ref{sec3}, we have concluded that in order to satisfy all hypotheses and all theoretical results, it is sufficient that the axis of symmetry of the magnetic dipole and the rotation axis can move one with respect to the other, which is never the case with a development in spherical harmonics (schematically illustrated in Figure \ref{fig:02}).
	
	Section \ref{sec4} has presented tests of our hypotheses on observations and other data. In  Figure \ref{fig:03} we have shown that sea-level variations in Brest, variations of declination in Chambon-la-Foret and Markowitz-Stoyko polar drift were essentially the same, except for the phase of declination, that we interpret as resulting from the nature of core-mantle coupling. We know from other tide-gauges \cite{Nakada2005,LeMouel2021}, that the Brest gauge is representative of most northern hemisphere gauges. \cite{Jault1991} discussed the correlation and the model that would link geomagnetic secular variations and polar drift; we extend the correlation and the model to sea-level. 
	
	Next, we have focused on annual and semi-annual oscillations, as did previously \blue{Jault and Le Mouël} (\cite{Jault1991}), here again bringing in the information carried by sea-level. All these cycles, regardless of whether they correspond to geomagnetic field components (X,Y,Z), length of day or sea-level, are in phase or in phase opposition. This observation falsifies the hypothesis that the field would include seasonal variations (which invoked the \blue{Russel and McPherron} \cite{Russell1973} effect), unless the effect could also explain the presence of the same cycles in the sea-level and length of day variations.
	
	We have calculated the ratios of mean amplitude of variations in these cycles of magnetic field components to associated tide gauges. Independent of local and regional geography and topography, we find these ratios to be quasi constant at $\sim$8mm/nT. This also agrees with the concept of a constant field (relation \ref{eq:08a}).
	
	We have next come to testing the famous Schwabe $\sim$11 yr cycle. Paradoxically it is rather weak in geomagnetic field components, and quite strong in magnetic indices, such as aa. In a parallel way, the cycle is weak in polar motion yet it is one of the main components of length of day. This is readily understood in \Laplace ’s paradigm (\cite{Laplace1799}): polar motion and length of day describe the same phenomenon but differ by one order of derivation. In the same way, aa is also a derivative operator (difference between two antipodal observatories). We have shown that the 11yr component of the magnetic field is in phase with the corresponding component of polar motion, when the latter (\textit{lod}) has been integrated (Figure \ref{fig:07}). In a reciprocal way, polar motion is linked to the derivative of magnetic components.
	
	Our last test has been with the secular variation of the \textbf{IGRF}. Despite the caveat that a dynamic magnetic field derives from a vector potential, not a scalar potential, hence can in principle not be developed in spherical harmonics (Poisson, Gauss), regular analyses of magnetic field models have been produced at five year intervals for years since 1900. The secular variation of that field has been monotonous and decreasing, as far as the leading axial dipole field component $g_{1,0}$ is concerned. This behavior is parallel to that of polar motion ($m_{1},m_{2}$, see Figure \ref{fig:11a}) or also $\dot{m}_{3}$.
	
	These two lines of observations complement our series of tests of the validity (and foresight) of Poisson’s derivation of Maxwell’s equations, Larmor’s equation, the Liouville-Euler system, and in general of the distinction that must be made between electric and magnetic fields. The parallel behaviors of magnetism and rotational mechanics, illustrated by the Larmor formula, have been put to test with modern data (observations) with success, as we have endeavored to show in this paper. These results can be complemented by the analysis of the responses at a series of “commensurate periods” that have their origins in the ballet of (primarily) Jovian planets: these are the "forcing factors" not only of variations in sea-level, and a number of geophysical, atmospheric and heliophysical processes (\eg \cite{Scafetta2012,Stefani2016,Courtillot2021}) but also of Earth’s magnetic field as we have shown in this paper.
	
\bibliographystyle{ieeetr}
\bibliography{poisson}
\end{document}